\documentclass[sigconf]{acmart}

\usepackage{hyperref}
\usepackage[ruled, linesnumbered, vlined, commentsnumbered]{algorithm2e}
\usepackage{amsmath,blkarray,booktabs}
\usepackage{xcolor,colortbl}
\usepackage{multirow}
\usepackage{subcaption} 
\usepackage{tcolorbox}
\usepackage{csquotes}
\usepackage{tikz,pgfplots}
\usepackage{standalone}
\usepackage{pgfplots}
\usepackage{graphicx}
\usepackage{balance}
\usepackage{pifont}

\usepackage{color}

{\authoralign{#1}
\ifblank{#2}
   {\def\shadequoteauthor{}\def\yshift{-2ex}\def\quotefill{\hfill}}
   {\def\shadequoteauthor{\par\authorfill\shadedauthorformat{#2}}\def\yshift{1ex}}
\begin{snugshade}\begin{quote}\openquote}
{\shadequoteauthor\quotefill\closequote{\yshift}\end{quote}\end{snugshade}}

\definecolor{one}{HTML}{2b7bba}

\newcommand{\vect}[1]{\overline{\boldsymbol{#1}}}

\newcommand{\changed}[1]{\textcolor{black}{#1}}

\definecolor{beaublue}{rgb}{0.74, 0.83, 0.9}

\newtcolorbox{quotebox}{colback=beaublue,boxrule=0.4pt,colframe=black,fonttitle=\bfseries,top=2pt,bottom=2pt}

\newcommand{\Model}{\texttt{DHDA}} 

\newcommand{\CART}{\texttt{CART}}

\DeclareMathAlphabet\mathbfcal{OMS}{cmsy}{b}{n}

\AtBeginDocument{%
  }

\copyrightyear{2026}
\acmYear{2026}
\setcopyright{rightsretained}
\acmConference[ICSE '26]{2026 IEEE/ACM 48th International Conference on Software Engineering}{April 12--18, 2026}{Rio de Janeiro, Brazil}
\acmBooktitle{2026 IEEE/ACM 48th International Conference on Software Engineering (ICSE '26), April 12--18, 2026, Rio de Janeiro, Brazil}
\acmPrice{}
\acmDOI{10.1145/3744916.3764542}
\acmISBN{979-8-4007-2025-3/26/04}

\begin{document}


\title{Dually Hierarchical Drift Adaptation for Online Configuration Performance Learning}

\author{Zezhen Xiang}
\email{zz.xiang.work@gmail.com}
\authornote{Zezhen Xiang is also supervised in the IDEAS Lab.}
\affiliation{%
  \institution{School of Computer Science and Engineering\\University of Electronic Science and Technology of China}
  \city{Chengdu}
  \country{China}
}

\author{Jingzhi Gong}
\email{j.gong@leeds.ac.uk}
\affiliation{%
  \institution{School of Computer Science\\University of Leeds}
  \city{Leeds}
  \country{UK}
}

\author{Tao Chen}
\email{t.chen@bham.ac.uk}
\authornote{Tao Chen is the corresponding author.}
\affiliation{%
  \institution{IDEAS Lab\\School of Computer Science\\University of Birmingham}
  \city{Birmingham}
  \country{UK}
}

\renewcommand{\shortauthors}{Xiang et al.}

\begin{abstract}

Modern configurable software systems need to learn models that correlate configuration and performance. However, when the system operates in dynamic environments, the workload variations, hardware changes, and system updates will inevitably introduce \textit{concept drifts} at different levels{---\textit{global drifts}, which reshape the performance landscape of the entire configuration space; and \textit{local drifts}, which only affect certain sub-regions of that space}. As such, existing offline and transfer learning approaches can struggle to adapt to these implicit and unpredictable changes in real-time, rendering configuration performance learning challenging. To address this, we propose \Model, an online configuration performance learning framework designed to capture and adapt to these drifts at different levels. The key idea is that \Model~adapts to both the local and global drifts using dually hierarchical adaptation: at the upper level, we redivide the data into different divisions, within each of which the local model is retrained, to handle global drifts only when necessary. At the lower level, the local models of the divisions can detect local drifts and adapt themselves asynchronously. To balance responsiveness and efficiency, \Model~combines incremental updates with periodic full retraining to minimize redundant computation when no drifts are detected. Through evaluating eight software systems and against state-of-the-art approaches, we show that \Model~achieves considerably better accuracy and can effectively adapt to drifts with up to \changed{$2\times$} improvements, while incurring reasonable overhead and is able to improve different local models in handling concept drift.

\end{abstract}



\maketitle

\section{Introduction}
\label{sec:introduction}
Configurations are essential elements that impact the performance of modern software systems, e.g., latency and throughput~\cite{DBLP:journals/tse/SayaghKAP20, DBLP:journals/tosem/ChenL23a}. {As a few examples, for video encoders such as \textsc{x264}, inappropriate configurations might cause $10\times$ performance degradation~\cite{DBLP:journals/jss/LesoilABJ23}; configuring the \texttt{query\_cache\_type} setting in \textsc{MySQL} can result in a performance boost greater than 11 times for specific workloads~\cite{DBLP:conf/cloud/ZhuLGBMLSY17}.}

Since how configurations can affect performance is complex~\cite{DBLP:journals/tse/SayaghKAP20}, a crucial task therein is to model the correlation between configurations and performance~\cite{DBLP:conf/esem/HanY16, chen2025accuracy}. Much work has been proposed, including white-box analytical models~\cite{DBLP:conf/icse/WeberAS21, DBLP:conf/wosp/HanYP21}, learning-based approaches~\cite{DBLP:conf/icse/HaZ19, DBLP:journals/tse/ChenB17, DBLP:journals/tosem/ChengGZ23, DBLP:journals/ese/GuoYSASVCWY18, DBLP:journals/sqj/SiegmundRKKAS12}, and hybrid ones~\cite{DBLP:journals/corr/DidonaR14}. Recent research in this field has been focusing on learning in multiple environments, such as workloads, hardware, or versions~\cite{DBLP:journals/tse/KrishnaNJM21, DBLP:journals/pacmse/Gong024}. This makes sense, as real-world software systems are inevitably required to run under diverse conditions~\cite{muhlbauer2023analysing, DBLP:conf/icpp/MadireddyBCLLRS19}. {Therefore, it is essential to ensure that configuration performance models trained in one environment remain effective when the system operates in another, previously unseen environment.}

However, an unaddressed limitation of existing work is that they mostly assume offline learning/modeling, i.e., models are trained/built under historical environments using a fixed set of sample data and then applied to predict performance for the same or new environment, after which no further updates are considered. This ignores the true nature of running configurable systems: we do not often have a fixed set of configuration data, but those samples are collected continuously in an uncertain and dynamic manner~\cite{DBLP:conf/kbse/ZhangGBC15}, especially when the measurement of configurations is highly expensive~\cite{DBLP:conf/kbse/SarkarGSAC15, DBLP:journals/ese/GuoYSASVCWY18, DBLP:conf/sigsoft/JamshidiVKS18}. For example, when self-adapting the configurable system at runtime~\cite{DBLP:journals/tsc/ChenB17, DBLP:conf/wcre/Chen22}, a model, which informs planning decisions, can be continuously updated as measured data is collected for system adaptation under time-varying workloads. Yet, in the other case, profiling the system at deployment time might require changing the version and/or hardware, from which new data is collected, and the model can be updated to expedite the relevant testing~\cite{DBLP:conf/icse/MaChen25}. In those cases, the produced model should also be available anytime. This means that current approaches will struggle to adapt to the continuously changing and evolving characteristics of configuration performance over time, in which the concept to be learned is time-varying---a general phenomenon known as \emph{concept drift} \cite{DBLP:journals/tkde/LuLDGGZ19, DBLP:journals/sadm/DriesR09}. For example, an empirical study on the performance problems in configurable file systems \cite{DBLP:conf/fast/KasickTGN10} reveals that:  

\begin{displayquote}
\textit{``Anomalies like disk-hogs and packet-loss from concept drift disrupt software performance symmetry, underscoring the need for adaptive algorithms.''} 
\end{displayquote}

The above calls for an online configuration performance modeling approach as opposed to the current mainstream ``offline efforts''~\cite{DBLP:conf/icse/Chen19b}. Yet, learning configuration performance online is challenging due to the interplay between uncertain concept drifts and high sparsity of configuration data, where both the configuration options and samples can have radically different impacts on the performance~\cite{DBLP:conf/icpp/MadireddyBCLLRS19}.

While general online learning approaches exist~\cite{DBLP:conf/icpp/MadireddyBCLLRS19, DBLP:journals/tkde/LuLDGGZ19, DBLP:journals/csur/GamaZBPB14}, they do not fit well with our case because (1) they often fail to handle the sparsity of configuration data which harms their accuracy \cite{DBLP:conf/icpp/MadireddyBCLLRS19}. (2) They do not consider the interplay between concept drift and data sparsity, such that the concept drifts might only occur locally in the configuration space \cite{DBLP:journals/tse/GongCB25}. (3) These approaches rely on costly full retraining or ignore the expensive sample measurement of configurations \cite{DBLP:conf/icpp/MadireddyBCLLRS19}.

To fill the above gaps, in this paper, we present Dually Hierarchical Drift Adaptation, dubbed \Model, for learning configuration performance online. \Model~relies on \texttt{DaL}~\cite{DBLP:conf/sigsoft/Gong023}, a recent offline approach that handles sparsity in configuration data by building multiple local models for divided areas of the data, but significantly extends it for online learning. The unique property of \Model~is that it adapts to concept drift in a dually hierarchical way: {\emph{global drifts}} might be detected at the upper level, which triggers a full-scale adaptation of all local models. The lower level, i.e., at each local model, detects and adapts to {\emph{local drifts}} asynchronously and locally. In a nutshell, our key contributions include:

\begin{itemize}  

\item An upper level drift adaptation that monitors the states of all local models, then adapts to global drifts only when it is necessary.  

\item A lower level drift adaptation, through an extended drift detector, that only detects and adapts to local drifts that are happening in the local area of the configuration data, making drift adaptation more asynchronous and effective.


\item To maintain local models under normal conditions, we design a hybrid model maintenance mechanism that relies on both incremental updates and full retraining for consolidating the models in the absence of concept drifts.  
 
\item We evaluate \Model~by comparing it with five baseline/state-of-the-art approaches and a diverse set of local models, together with in-depth ablation analysis.

\end{itemize}

The results demonstrate that \Model~achieves the best accuracy on 75\% cases (6/8) with statistical significance and up to $2\times$ improvements. The model adaptation times are also highly acceptable, in a matter of seconds. All data and source code can be found at our repository: \texttt{\textcolor{blue}{\href{https://github.com/ideas-labo/dhda}{https://github.com/ideas-labo/dhda}}}.

The remainder of this paper is organized as follows. Section~\ref{sec:pre} presents the problem definitions and characteristics in online software performance learning. Section~\ref{sec:framework} introduces \Model~, detailing its key components. Section~\ref{sec:experiment_setup} describes the experimental setup, while Section~\ref{sec:evaluation} evaluates \Model~comprehensively. Section~\ref{sec:discussion} discusses key findings, whether \Model~is practically useful, and threats to validity. Section~\ref{sec:related_work} reviews related work, and Section~\ref{sec:conclusion} concludes the paper.

\section{Background}
\label{sec:pre}

\subsection{Configuration Performance Learning via \texttt{DaL}}
\label{sec:dal}

Without loss of generality, configuration performance learning seeks to build a function $f$ such that:
\begin{equation}
    \mathcal{P} = f(\mathbfcal{S})\text{, } \mathcal{P}\in\mathbb{R}
    \label{eq:prob}
\end{equation}
whereby $\mathbfcal{S}=\{\vect{s}_1,\vect{s}_2,...,\vect{s}_k\}$ is a set of training samples such that $\vect{s}_k=\{\vect{x}_k,p_k\}$. $\vect{x}_k = \{x_{1,k}, x_{2,k}, \dots, x_{n,k}\}$ is a configuration and $x_i$ is the value of the $i$th configuration option for sample $k$, which might be binary, categorical or enumerative. $\mathcal{P}$ is a performance metric, and the value for sample $k$ is $p_k$ (e.g., latency or throughput), measured under an environment.

Many solutions have been proposed to learn such a $f$~\cite{DBLP:conf/icse/HaZ19, DBLP:conf/kbse/JamshidiSVKPA17, DBLP:conf/icse/0003XC021, DBLP:journals/tse/GongCB25, DBLP:conf/sigsoft/Gong023}. More recently, one of the successful attempts is \texttt{DaL}~\cite{DBLP:conf/sigsoft/Gong023, DBLP:journals/tse/GongCB25}, a framework that caters for the \textbf{\textit{feature}} and \textbf{\textit{sample sparsity}} in configuration data. \texttt{DaL} reformulates Equation~\ref{eq:prob} from a regression problem to a mixed two-level problem such that the upper level is classification while the lower level is regression:
\begin{equation}
    \mathbfcal{D} = g(\mathbfcal{S}) 
    \label{eq:dal_1}
\end{equation}
\begin{equation}
    \forall D_i \in \mathbfcal{D}\text{: } \mathcal{P} = f(D_i)\text{, } \mathcal{P}\in\mathbb{R}
    \label{eq:dal_2}
\end{equation}
where $D_i$ is a division that consists of proportional samples from the training dataset. In a nutshell, \texttt{DaL} works as follows:

\begin{enumerate}
    \item \textbf{Dividing:} Divide the training samples into diverse divisions $\mathbfcal{D}$ using {Classification and Regression Tree (\texttt{CART})~\cite{lewis2000introduction}} with a depth $d$ (building function $g$). The samples-division mappings are used to train a Random Forest classifier.
    \item \textbf{Training:} Train a dedicated local model for each division $D_i$, containing samples extracted at the $d$ depth from the trained \texttt{CART} (building function $f$). 
    \item \textbf{Predicting:} Leverage the Random Forest classifier to assign a newly coming configuration into the right model for prediction (using functions $g$ and $f$).
\end{enumerate}
\texttt{DaL}, as a generic framework, has been shown to be superior to many state-of-the-art approaches. The unique property is that \texttt{DaL} can be paired with an arbitrary approach as the local model, hence improving its ability to handle sample sparsity to consolidate accuracy. As such, we leverage \texttt{DaL} as the foundation in this work for online scenarios.

\subsection{Online Configuration Performance Learning}

To cater for the dynamically evolving environment under uncertainty, i.e., workloads, versions, and hardware, of a configurable system, the configuration performance learning can be best modeled as learning online. Formally, this means
\begin{equation}
    \mathcal{P} = f_t(\mathbfcal{S}_t)\text{, } \langle \{\vect{x}_1,p_1\}, \{\vect{x}_2,p_2\},...,\{\vect{x}_m,p_m\} \rangle \rightarrow \mathbfcal{S}_t
    \label{eq:new_prob}
\end{equation}
Similar to the offline scenario in Equation~\ref{eq:prob}, online configuration performance learning also seeks to build the function $f$. However, since the new configuration samples are collected continuously as a data stream in a dynamic and uncertain manner: at time $t$, the training sample is enriched in time-varying paces with one or more newly measured samples ($\mathbfcal{S}_t$) and hence the function $f_t$ needs to be updated. This is the root cause of concept drift~\cite{DBLP:journals/tkde/LuLDGGZ19, DBLP:conf/fast/KasickTGN10, DBLP:journals/jksucis/AgrahariS22}, where the concept of correlation between a performance metric and configuration can change over time, depending on the time-varying environment. By online, we refer to either system design time or runtime, in which the former often occurs in Continuous Integration/DevOps such that the system is continuously profiled under different environments~\cite{muhlbauer2023analysing, DBLP:conf/kbse/JamshidiSVKPA17} while the latter means runtime monitoring/sampling of a system (or its digital twin)~\cite{DBLP:conf/wcre/Chen22, DBLP:journals/tosem/ChenLBY18}. 

It is worth noting that online configuration performance learning differs from the existing multi-tasks/transfer configuration performance learning~\cite{DBLP:journals/tse/KrishnaNJM21, DBLP:journals/pacmse/Gong024}. This is because, in the latter, the different environments can be modeled explicitly and with certainty, i.e., we know when and how the environment changes. However, in online configuration performance learning, the environment changes implicitly and in an uncertain way, e.g., a system might run under a workload $A$ for some time before experiencing another workload $B$, after which workload $A$ reoccur again, or the system might require to conduct a live migration to a (virtual) machine with different hardware. Therefore, we do not know when and at what scale those changes occur.

Yet, most existing works and \texttt{DaL} have not been designed for such an online learning scenario, which is what we seek to address in this work.

\subsection{Problem Characteristics}
\label{subsec:characteristics}

\begin{figure}[!t]
\centering
\footnotesize



\begin{subfigure}{0.5\columnwidth}
  \centering
  \includegraphics[width=\linewidth]{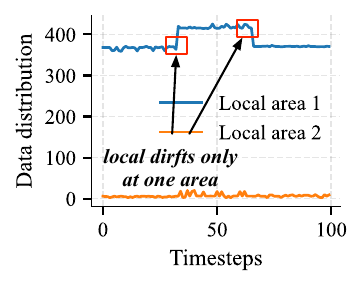} 
  ~\vspace{-0.2cm}
  \caption{Local drifts in \textsc{NGINX}}
\end{subfigure}
~\hfill
\begin{subfigure}{0.5\columnwidth}
  \centering
    \includegraphics[width=\linewidth]{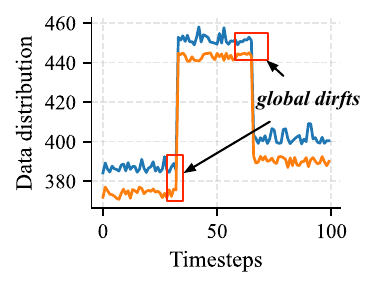} 
  ~\vspace{-0.2cm}
  \caption{Global drifts in \textsc{SQLite}}
\end{subfigure}

\begin{subfigure}{0.49\columnwidth}
  \centering
  ~\vspace{-0.1cm}
  \includegraphics[width=\linewidth]{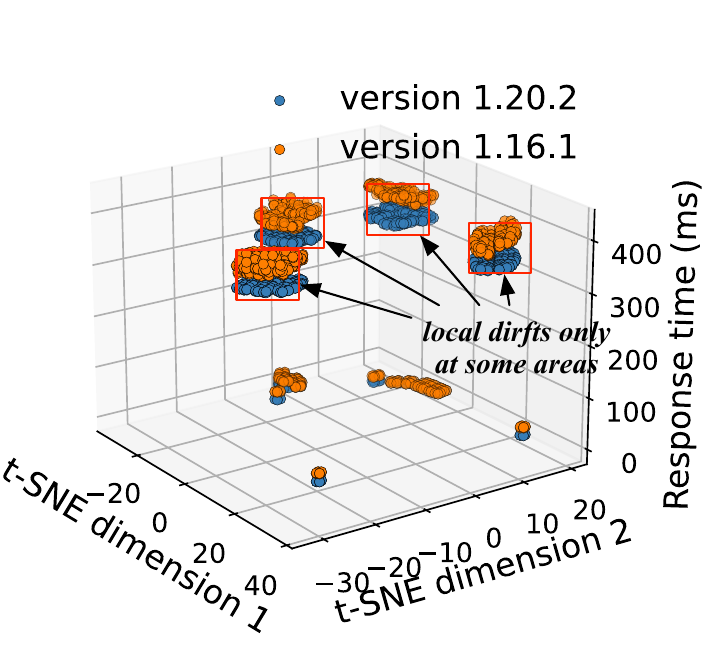}
  \caption{Landscape for \textsc{NGINX}}
\end{subfigure}
~\hfill
\begin{subfigure}{0.49\columnwidth}
  \centering
   ~\vspace{-0.12cm}
  \includegraphics[width=\linewidth]{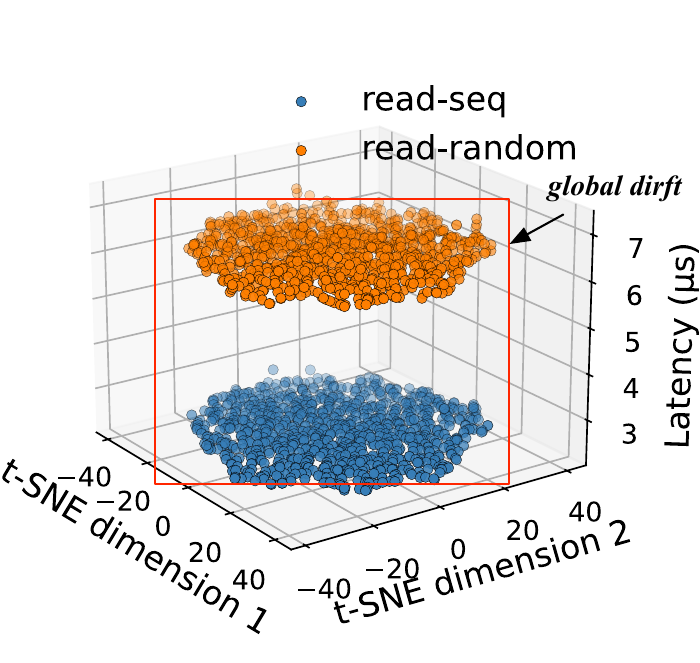}
  \caption{Landscape for \textsc{SQLite}}
\end{subfigure}

  \caption{Examples of local and global drifts for systems. {(c) and (d) apply t-SNE~\cite{van2008visualizing} to project the high-dimensional configuration space onto two representative dimensions.}}
     \label{fig:drifts}
\end{figure}





To understand the challenges and characteristics of online configuration performance learning, we study the datasets collected from commonly used configurable systems and their workloads in prior studies~\cite{DBLP:journals/tse/KrishnaNJM21, DBLP:conf/kbse/JamshidiSVKPA17, DBLP:conf/icse/webertwins}. Figure~\ref{fig:drifts} shows the common examples of systems under continuous environment changes and we observe:

\begin{itemize}
    \item \textbf{Characteristic 1:} It is uncertain when the concept drift occurs due to time-varying environments and sample measurements (Figures~\ref{fig:drifts}a and~\ref{fig:drifts}b).

    \item \textbf{Characteristic 2:} Because of configuration data sparsity, concept drifts might occur globally (Figures~\ref{fig:drifts}b and d) or only in local areas (Figures~\ref{fig:drifts}a and c). For example, we see that for system \textsc{NGINX}, the discrepancy in the landscape only occurs in local areas when the version changes. This can also be reflected in the actual data stream, where concept drift might occur in one area of the configuration sample space but not the other.

\end{itemize}

In addition, the nature of configurable systems and their quality assurance pipeline implies specific requirements on the overhead of online configuration performance learning:

\begin{itemize}
    \item \textbf{Characteristic 3:} Due to costly measurements~\cite{DBLP:conf/sigsoft/JamshidiVKS18} and the dynamic nature of system performance, the learning needs to be efficient for real-time computation, allowing ``anytime'' property for model exploitation.

\end{itemize}

Therefore, simply retraining \texttt{DaL} (or other offline learning approaches) in an ad-hoc manner would certainly fail to handle the above challenges, which, as we will show, can lead to devastating outcomes. In contrast, the general online learning approaches~\cite{DBLP:conf/esann/GomesBFB18, DBLP:conf/icdm/GomesRB19} can fail to handle the specific sparsity issue in configuration data, and they are not tailored to the patterns of concept drifts that are most common for online configuration performance learning (\textbf{Characteristics 2}).

\section{Online Learning with \Model}
\label{sec:framework}


Drawing on the characteristics, we propose \Model~for online configuration performance learning leveraging \texttt{DaL}. Like all online learning approaches, \Model~maintains a sliding window $\mathbfcal{W}$ that contains $q$ incrementally measured configuration samples. As such, \Model~changes the problem at timestamp $t$ as:
\begin{equation}
    \mathbfcal{D}_{t} = g_{t}(\mathbfcal{W})
\end{equation}
\begin{equation}
    \forall D_{i, t} \in \mathbfcal{D}_{t}\text{: } \mathcal{P} = f_{t}(D_{i,t})\text{, } \mathcal{P} \in \mathbb{R}
\end{equation}
Here, each local model would also have their own sliding window $\mathbfcal{W}_i$. The key idea is that although updating the \texttt{CART} ($g_{t}$) and the changes of the local models ($f_{t}$) based on their divisions of samples $D_{i,t}$ happen at the same timestamps, the actual samples used and the mechanism of model updates could differ among different local models, leading to different extents/aspects of local model change. This dually hierarchical adaptation in handling concept drifts is our key to fit the characteristics observed from Section~\ref{subsec:characteristics}. 


\begin{figure}[t!]
  \centering
  \includegraphics[width=\columnwidth]{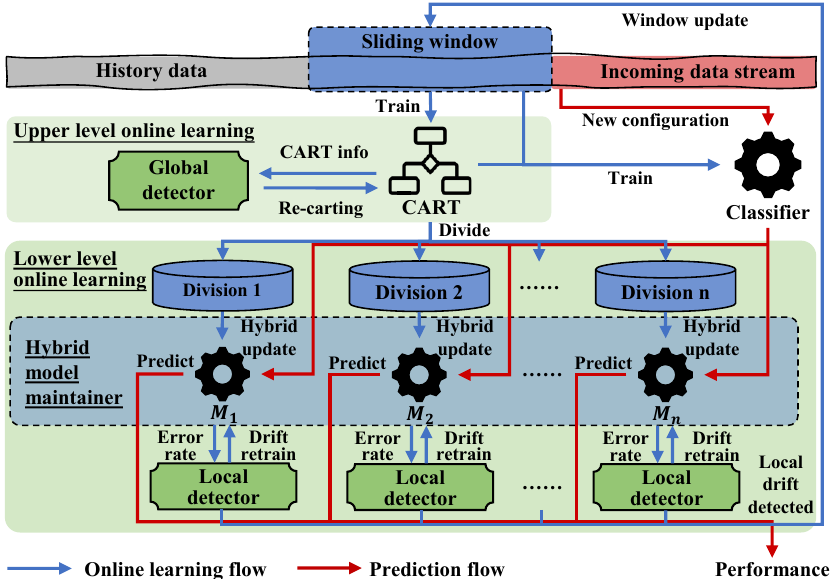}
  \caption{The general overview of \Model.}
  \label{fig:structure}
\end{figure}

Figure~\ref{fig:structure} and Algorithm~\ref{alg:learning-code} show the workflow of \Model~with the following key components for learning a continuously measured stream of configuration data online:

\begin{algorithm}[t!]
	\DontPrintSemicolon
	\footnotesize
	
	\caption{Pseudo code of online learning in \Model}
	\label{alg:learning-code}
	\KwIn{The depth $d$; and number of timesteps of retaining $\alpha$; Sliding window at upper level $\mathbfcal{W}$; windows at lower level for each local model $\mathbfcal{W}_{i}$; a new batch of samples at timestep $t$ $\mathbfcal{S}_t$}

	
      
      

     \For{every timestep $t$}  
      {
         $\mathbfcal{W} = \mathbfcal{W} \cup \mathbfcal{S}_t$\\
         $\mathbfcal{C}_t\leftarrow$ \textsc{trainCART($\mathbfcal{W}$)}\\
       
         $\mathbfcal{D}\leftarrow$ extract all divisions from $\mathbfcal{C}_t$ under $d$\\
           $\mathbfcal{W}_1,\mathbfcal{W}_2,...,\mathbfcal{W}_n\leftarrow$ assign the new samples into the windows of the local model based on $\mathbfcal{D}$\\
	$\mathbfcal{F}\leftarrow$ \textsc{trainRandomForest($\mathbfcal{D}$)}\\

  \tcc{\textcolor{blue}{the initialization at the first timestep only.}}
 \If{$\mathbfcal{C}_c = \emptyset$}{

 $\mathbfcal{C}_c = \mathbfcal{C}_t$\\
      \For{$\forall D_i \in \mathbfcal{D}$}  
      {
       $\mathbfcal{M}\leftarrow$ \textsc{trainLocalModel($D_i$)}\\
    
      }

 }
      
       \tcc{\textcolor{blue}{upper level online learning.}}
         \If{$G(\mathbfcal{C}_t) - G(\mathbfcal{C}_c) > \epsilon$}
	   {


clear $\mathbfcal{M}$ and all local windows\\

    \For{$\forall D_i \in \mathbfcal{D}$}  
      {
       $\mathbfcal{M}\leftarrow$ \textsc{trainLocalModel($D_i$)}\\
    
      }
      	
               $\mathbfcal{C}_c = \mathbfcal{C}_t$\\
          }

          \Else {

 \For{$\forall \mathbfcal{M}_{i,c} \in \mathbfcal{M}$}  
 {

 \tcc{\textcolor{blue}{lower level online learning.}}
      \If{ADWIN detects drifts on $\mathbfcal{M}_{i,c}$ over $\mathbfcal{W}_{i}$}  
      {
      
        discard the out-of-date samples in $\mathbfcal{W}_{i}$\\

        $\mathbfcal{M}_{i,t}=$ \textsc{trainLocalModel($\mathbfcal{W}_i$)}\\
          
      } \Else {
       \tcc{\textcolor{blue}{hybrid model maintaining.}}
          $t'\leftarrow$ count the number of timesteps that have new data samples assigned to $\mathbfcal{W}_{i}$ since last drift\\
          \If{$t' \mod \alpha \neq 0$} {
           $\mathbfcal{W'}_i\leftarrow$ extract the new data from $\mathbfcal{W}_{i}$\\
          $\mathbfcal{M}_{i,t}=$ \textsc{updateLocalModel($\mathbfcal{M}_{i,c}$,$\mathbfcal{W'}_i$)}\\
          } \Else {
            $\mathbfcal{M}_{i,t}=$ \textsc{trainLocalModel($\mathbfcal{W}_i$)}\\
          }

      }
      replace $\mathbfcal{M}_{i,c}$ with $\mathbfcal{M}_{i,t}$\\
 
 }

      }
      }

\end{algorithm}

\begin{itemize}
    \item \textbf{Upper Level Online Learner:} At lines 12-17, we detect global drift at the upper level, and trigger the full-scale local models adaptation only if needed (\textbf{Characteristics 1-2}).
    \item \textbf{Lower Level Online Learner:} At the lower level, we detect local drift for the model of each division. Adaptation of different models is considered in isolation, as shown in lines 20-23 (\textbf{Characteristics 1-2}). 
    \item \textbf{Hybrid Model Maintainer:} At lines 25-32, during normal operation, we adopt a hybrid training mechanism that combines incremental updates and re-training to consolidate the local models (\textbf{Characteristic 3}).
  
\end{itemize}

In this way, global drifts occurring at the upper level would be detected and trigger a full-scale adaptation while local drifts happening in each individual local model can be detected and handled locally, without affecting the entire \texttt{DaL}. Local model adaptation can also benefit from hybrid adaptation.

Upon prediction, \Model~basically follows the procedure of \texttt{DaL} to assign the given configuration into the suitable division and its local model, which then predicts the performance (Algorithm~\ref{alg:pred-code}).

\begin{algorithm}[t!]
	\DontPrintSemicolon
	\footnotesize
	
	\caption{Pseudo code of prediction in \Model}
	\label{alg:pred-code}
	\KwIn{A new configuration $\mathbf{\overline{c}}$ to be predicted; a random forest $\mathbfcal{F}$, and the set of local models $\mathbfcal{M}$}
    \KwOut{The predicted performance of $\mathbf{\overline{c}}$}

    $D_i=$  \textsc{predict($\mathbfcal{F}$,$\mathbf{\overline{c}}$)}\\
    $\mathcal{M}=$ get the model from $\mathbfcal{M}$ that corresponds to the predicted division $D_i$\\

    \Return \textsc{predict($\mathcal{M}$,$\mathbf{\overline{c}}$)}\\
	
\end{algorithm}

\subsection{Learning Upper Level Online (Global Drifts)} 

The natural design of \texttt{CART} in \texttt{DaL} is that it determines the produced divisions, which then affect the local models trained. Therefore, any change at the upper level would require sustainable retraining of all the local models at the lower level (\textbf{Characteristic 2}).

\subsubsection{Drift Detection}

To avoid excessive full scale adaptation and meet \textbf{Characteristic 3}, as a batch of new samples arrives at time $t$, we compute the additional information entropy, quantified by Gini importance~\cite{DBLP:journals/ml/Breiman01}, when using all new data samples arrived in $\mathbfcal{W}$ so far to retrain the \texttt{CART} (since retraining \texttt{CART} using $\mathbfcal{W}$ is cheap). The result of this is compared against an extended Hoeffding Bound~\cite{DBLP:phd/basesearch/Tao11}:
\begin{equation}  
\epsilon = \sqrt{\frac{\ln({1 \over \delta})}{2h}}  \text{ s.t., } h={L \over {\sum^L_{i=1}{1\over{n_i}}}}
\end{equation}  
$\epsilon$ is the threshold that caps the independent random variables that deviate from their expected values by more than a certain amount. $\delta$ is the significance level ($\delta=0.05$ in our work). In the original Hoeffding Bound, $h$ denotes the total number of samples, but it is ill-fitted to our case as (1) it has not considered the contributions of data belonging to individual divisions at depth $d$ in \texttt{DaL} and (2) the data/drifts in different divisions can be highly imbalanced, e.g., most newly arriving samples might belong to one division more than the others. Therefore, we set $h$ as the harmonic mean of sample numbers from all divisions, where $n_i$ is the number of samples belongs to the $i$th division; $L$ is the number of divisions at depth $d$ and $L \in [d+1,2^d]$~\cite{DBLP:journals/tse/GongCB25}. For example, if $d=1$, then certainly we only have two divisions.

When all new samples cumulated so far lead to additional information entropy beyond the extended Hoeffding Bound, it means that those new data samples should cause sustainable, often sudden, changes of the divisions, and hence updating all local models is necessary (\textbf{Characteristic 1-2}).

Specifically, \Model~adapts a drift detector that works as follows:

\begin{enumerate}
    \item Use all samples in $\mathbfcal{W}$ to retain a new \texttt{CART} $\mathbfcal{C}_t$.
    \item Calculate the deviation between the total Gini importance~\cite{DBLP:journals/ml/Breiman01} $G$ of $\mathbfcal{C}_t$ and that of the current \texttt{CART} $\mathbfcal{C}_c$:
    \begin{equation}  
\Delta G = G(\mathbfcal{C}_t) - G(\mathbfcal{C}_c)
\end{equation}  
     \item If $\Delta G > \epsilon$, then we detect a drift at the upper level that needs to be handled.
\end{enumerate}

\subsubsection{Drift Mitigation}

When a drift at the upper level is detected, \Model~replaces $\mathbfcal{C}_c$ with $\mathbfcal{C}_{t}$ and performing a ``redividing'' process (updating $g_t$), producing a new set of divisions and training all local models for the divisions. Here, every local model needs to be completely retrained, as drift detected at the upper level often implies a sudden and global drift. The Random Forest classifier in \texttt{DaL} is retrained with the new data samples regardless of the global drift since the local models need to be adapted under normal situations (see Section~\ref{subsec:hybrid_model_update}).

\subsection{Learning Lower Level Online (Local Drifts)} 
At the lower level, each division is trained with a local model, hence the localized drifts are detected and handled within the division.

\subsubsection{Drift Detection}

To detect local drifts, we use the ADWIN \cite{DBLP:conf/sdm/BifetG07,gama2004learning}---a widely used solution---as the drift detector to monitor the extents of local drifts (\textbf{Characteristic 1-2}). However, ADWIN only detects the performance change of a model regardless of whether the change is beneficial. This is problematic since if the error is generally decreasing, then it often means that the model is gradually adapting to the concept rather than a concept drift. As such, we pair a trace filter~\cite{DBLP:conf/dis/GunasekaraGBP22} that compares the traces of two time-series and determines whether they are increasing or decreasing; only the increasing one is of concern.

In a nutshell, at timestep $t$ (if no global drift detected), \Model~uses a retained \texttt{CART} to assign the new samples into the window $\mathbfcal{W}_i$ for the $i$th current model $\mathbfcal{M}_{i,c}$ adapted at timestep $c$. ADWIN, paired with the trace filter, splits $\mathbfcal{W}_i$ into two parts that deviate the most in terms of model performance, i.e., the order samples ($\mathbfcal{W}_{i,l}$) and the newer samples ($\mathbfcal{W}_{i,r}$), and monitors the performance of $\mathbfcal{M}_{i,c}$ in two states (with default settings):
\begin{itemize}  
    \item \textbf{Warning state:} If the overall degradation of accuracy from $\mathbfcal{W}_{i,l}$ to $\mathbfcal{W}_{i,r}$, as identified by the filter, is more than a relaxed threshold (90\%), then we flag a warning.
    \item \textbf{Drift state:} If there is already a previously flagged warning state and the overall degradation of accuracy from $\mathbfcal{W}_{i,l}$ to $\mathbfcal{W}_{i,r}$, as identified by the filter, is more than a stricter threshold (99\%), then we detect a drift.
\end{itemize}

To measure accuracy, we adopt the widely used, scale-invariant Mean Absolute Percentage Error (MAPE) in \Model: $\frac{1}{N} \sum_{i=1}^{N} \frac{|y_i - \hat{y}_i|}{y_i}$, where $y_i$ and $\hat{y}_i$ are the actual and predicted performance, respectively, for a total of $N$ samples.

ADWIN also helps \Model~to determine which samples in the sliding window $\mathbfcal{W}_i$ should be discarded when there is a drift: any samples between the latest warning state and the drift are considered as timely data, while any previous data samples in $\mathbfcal{W}_i$ is the out-of-date data and can be discarded---this is the key to keep the model always learning the ``right'' concepts. The same samples in the global window $\mathbfcal{W}$ are also removed.

\subsubsection{Drift Mitigation}

Once a local drift is detected for $\mathbfcal{M}_{i,c}$, \Model~resets the detector, discards out-of-date data, and adapts $\mathbfcal{M}_{i,c}$ using the remaining data in $\mathbfcal{W}_i$. The model adaptation follows the standard retraining process. We would like to stress that the local drift at different local models can be detected and mitigated asynchronously, hence fitting \textbf{Characteristics 2}.

\subsection{Hybrid Model Maintaining}  
\label{subsec:hybrid_model_update}  

The above describes the need for model adaptation when concept drifts occur. In addition, we also observe that even in a normal situation, newly arriving data can be helpful in improving the performance of configuration performance prediction, even though the concept remains unchanged. As such, \Model~also maintains each local model under normal operation using new samples in $\mathbfcal{W}_i$ without discarding any data. However, given \textbf{Characteristic 3}, adapting the model by frequent retraining is not ideal. To that end, at each timestep, \Model~applies a hybrid adaptation mechanism that adapts the model using both incremental update and retraining complementarily. 

{Specifically, for the incremental update, we adapt the current model $\mathbfcal{M}_{i,k}$ whose weights/structure are denoted by $\theta$, adjusting them with the newly arrived samples $\mathbfcal{W}'_i$, thus yielding the updated model $\mathbfcal{M}_{i,t}$ (the \textsc{updateLocalModel} function in Algorithm~\ref{alg:learning-code}):}
\begin{equation}  
\mathbfcal{M}_{j,k} + \theta \cdot f(\mathbfcal{W}'_i) \Rightarrow  \mathbfcal{M}_{i,t}
\end{equation} 
In contrast, for retraining, we use all samples in the local windows ($\mathbfcal{W}_i$) accumulated so far for building a new model:
     \begin{equation}  
f(\mathbfcal{W}_i) \Rightarrow \mathbfcal{M}_{i,t} 
\end{equation} 

\begin{figure}[!t]
\centering
\footnotesize

  \centering
  \includegraphics[width=0.9\linewidth]{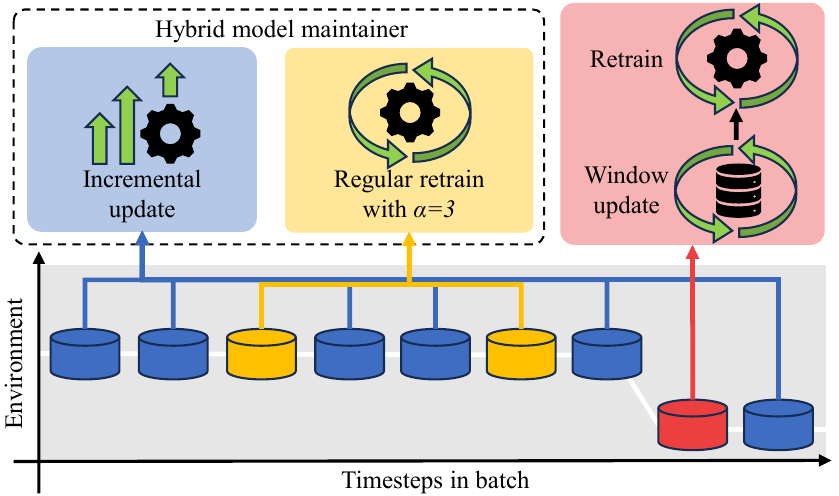} 
  \caption{Hybrid model maintenance in \Model.}

     \label{fig:hybrid_update_example}
\end{figure}

As in Figure~\ref{fig:hybrid_update_example}, whenever a batch of new data samples arrives, we adapt the local models. The decision as to whether the model is adapted following incremental update (based on the current model) or completely retrained is controlled by a parameter $\alpha$, which is the number of timesteps where there are new samples assigned to the window of the local model before the next retraining cycle. It is worth noting that this differs from a full-scale adaptation triggered by global drifts in several aspects:

\begin{itemize}
    \item Updating the local models under normal conditions can be done incrementally, as there is no concept drift. Full-scale adaptation, in contrast, requires retraining to handle the global drifts.
    \item Since there is no drift, there is no need to redivide the data samples.  
    \item When retraining is needed, the local models are retained asynchronously in the absence of concept drift, i.e., not all the models might be adapted at the same timestep depending on how the new data samples are assigned to their windows, since some local models might not have any new data for a timestep. In full-scale adaptation, all models need to be retained at the same time.
   
\end{itemize}

In this way, we enable local models to keep track of the new data information (which might not cause drifts) while minimizing the overhead incurred in model adaptation.

\section{Experiment Setup}
\label{sec:experiment_setup}

In this section, we describe the setup used to evaluate \Model~in online configuration performance learning scenarios. All experiments are conducted on a server with  Ubuntu, Intel(R) Xeon(R) Platinum, 40-core CPUs and 500GB RAM.



\subsection{Research Questions}
To comprehensively assess \Model, we seek to answer the following research questions (RQs):  
\begin{itemize}
    \item \textbf{RQ1:} How does \Model~compare to state-of-the-art approaches in terms of accuracy and computational cost?
    \item \textbf{RQ2:} How does \Model~impact different local models used for online performance learning?
    \item \textbf{RQ3:} How do the key designs in \Model~contribute? 
    \item \textbf{RQ4:} What is the sensitivity of \Model~to its parameter $\alpha$?
\end{itemize}

In all cases, unless otherwise stated, we pair \Model~with Random Forest as the default local model due to its high model adaptation efficiency for online learning.

\subsection{Subject Configurable Systems}  
\label{subsec:subject_systems}  

We rely on commonly used datasets from prior studies \cite{DBLP:conf/sigsoft/JamshidiVKS18, DBLP:conf/icse/webertwins, DBLP:conf/kbse/JamshidiSVKPA17, DBLP:journals/tse/KrishnaNJM21}, covering eight widely used configurable software systems across diverse domains, including deep learning frameworks, web servers, cloud computing tools, database management systems, and big data analyzers. They are selected based on their prevalence in state-of-the-art performance learning, the diverse behaviors in performance, and the presence of multiple environments. Table~\ref{tb:subject_systems} provides an overview of the subject systems.



\subsubsection{Noise and Bias Mitigation} 

The chosen datasets (systems and their environments) have been rigorously curated and assessed, ensuring that the evaluation remains reliable, unbiased, and comparable with prior research \cite{DBLP:conf/sigsoft/JamshidiVKS18, DBLP:conf/icse/webertwins, DBLP:conf/kbse/JamshidiSVKPA17, DBLP:journals/tse/KrishnaNJM21}. Specifically, the datasets have been collected with the following major strategies:

\begin{itemize}
    \item To reduce measurement noise, during dataset collection, each configuration was measured three times under the same conditions, and the mean execution time was stored.
    
    \item Random or specific sampling was applied in systems with large configuration spaces to ensure coverage of diverse performance behaviors while maintaining feasibility. For instance, in \textsc{SQLite}, 1000 configurations were sampled to capture variations in indexing and compression settings. 
    \item Environment diversity was explicitly considered to simulate different real-world scenarios\footnote{{The detailed environments for each system can be found at:\\ \href{https://github.com/ideas-labo/dhda/tree/main/drift_data}{\texttt{\textcolor{blue}{https://github.com/ideas-labo/dhda/tree/main/drift\_data}}}}.}. For example, \textsc{SQLite} was measured on query types covering sequential reads, random reads, sequential writes, and batch writes, which can change unexpectedly.
\end{itemize}

Indeed, the scale of the dataset might not be as large as some other software engineering problems~\cite{DBLP:conf/kbse/LiXCT20,DBLP:conf/icse/LiX0WT20}. The key reason is the expensive data collection for configurable systems: it has been reported that even for a single configuration, it can take up to 166 minutes to complete one measurement run~\cite{DBLP:conf/sigsoft/0001L21,DBLP:journals/corr/abs-2112-07303}, let alone the fact that each sample in the datasets is averaged from 5 repeats/measurements. As such, the datasets with up to $16,000$ samples per environment (e.g., up to 14 diverse workloads, 2 versions, and 2 hardware) are already promising. 

For the dimensions of options, we follow what has been collected/identified in the datasets~\cite{DBLP:conf/kbse/JamshidiSVKPA17,DBLP:conf/sigsoft/JamshidiVKS18,DBLP:journals/tse/KrishnaNJM21,DBLP:conf/icse/webertwins} and those options that are insensitive to performance have been removed. For example, a recent study~\cite{DBLP:conf/icse/LiangChen25} shows that \textsc{nginx} has $79$ options, but only $15\%$ are performance sensitive, which can be easily filtered out by reading the documentation/code. Therefore, we hope to avoid creating an artificially complex problem with dimensionality more than necessary. 

\begin{table}[t!]
\caption{Details of the subject systems. ($|\mathbfcal{B}|$/$|\mathbfcal{N}|$) denotes the number of binary/numerical options, $|\mathbfcal{C}|$ denotes the number of valid configurations per environment (full sample size), and $|\mathbfcal{E}|$ denotes the total number of environments in the data streams. $|\mathbfcal{H}|$, $|\mathbfcal{W}|$, and $|\mathbfcal{V}|$ respectively denotes the number of hardware, workload, and version considered.}
\vspace{-0.3cm}
\centering
\footnotesize
\begin{adjustbox}{width=\linewidth,center}
\setlength{\tabcolsep}{1mm}
\begin{tabular}{l|lcrrr|ccc}
\toprule

\multirow{2}{*}{\textbf{System}} & \multirow{2}{*}{\textbf{Domain}} & \multirow{2}{*}{\textbf{$|\mathbfcal{B}|$/$|\mathbfcal{N}|$}} & \multirow{2}{*}{\textbf{$|\mathbfcal{C}|$}} & \multirow{2}{*}{\textbf{$|\mathbfcal{E}|$}} & \multirow{2}{*}{\textbf{Ref.}} & \multicolumn{3}{c}{\textbf{Environments}} \\
 &  &  &  &  &  & \textbf{$|\mathbfcal{H}|$} & \textbf{$|\mathbfcal{W}|$} & \textbf{$|\mathbfcal{V}|$} \\ \hline
\textsc{SaC} & Cloud tool & 58/0 & 4999 & 10 & \cite{DBLP:conf/kbse/JamshidiSVKPA17} & 1 & 10 & 1 \\
\textsc{x264} & Video encoder & 16/0 & 2047 & 21 & \cite{DBLP:conf/kbse/JamshidiSVKPA17} & 6 & 3 & 2 \\
\textsc{Storm} & Big data analyzer & 1/11 & 2048 & 8 & \cite{DBLP:conf/sigsoft/JamshidiVKS18} & 1 & 4 & 1 \\
\textsc{SPEAR} & Audio editor & 14/0 & 16384 & 10 & \cite{DBLP:conf/kbse/JamshidiSVKPA17} & 3 & 4 & 2 \\
\textsc{SQLite} & Database & 14/0 & 1000 & 15 & \cite{DBLP:journals/tse/KrishnaNJM21} & 2 & 14 & 2 \\
\textsc{NGINX} & Web server & 16/0 & 1104 & 4 & \cite{DBLP:conf/icse/webertwins} & 1 & 1 & 4 \\
\textsc{ExaStencils} & Code generator & 8/4 & 4098 & 4 & \cite{DBLP:conf/icse/webertwins} & 1 & 4 & 1 \\
\textsc{DeepArch} & DNN tool & 12/0 & 4096 & 3 & \cite{DBLP:conf/sigsoft/JamshidiVKS18} & 3 & 1 & 1

\\
\bottomrule
\end{tabular}
\end{adjustbox}
\label{tb:subject_systems}
\end{table}

\subsection{Experiment Procedure with Data Streams}

{To emulate real-world concept drifts, we randomly mix the samples from different environments of a system (for each run) to form a data stream (e.g., system with $5000$ configurations and 12 environments would have $12\times5000=60000$ samples), since real-life cases are chaotic/uncertain: we do not know when a workload will change nor when we migrate the hardware. The steps are:}

\begin{enumerate}

\item  We randomly select the batches of samples (with 32 samples per batch) from the mixed workloads and sequentially feed them into a model for prediction\footnote{To ensure the realism of the experiment under our resources, we constrain the data samples from each environment to be $4000$.}. The data stream is built without replacement, i.e., the sample that has been fed will not be used again. In total, for a data stream with 3 environments, we have ${{3\times4000\over32}}=375$ timesteps, each with 32 newly arrived samples.


\item {By the end of timestep $t$, the model might be updated using data up to $t$ (all or partial); the updated model will predict all samples for the newly arrived batch at timestep $t+1$ and compute the accuracy/MAPE for 32 new samples}.

\item {The process is repeated for the next timestep---a standard procedure for online learning/concept drifts~\cite{DBLP:journals/csur/GamaZBPB14}.}

\end{enumerate}

As such, the evaluation always focuses on the generalization error: predictions on data not used for model updating/training.

\subsection{Metrics and Statistical Validation}  
\label{subsec:metrics_validation}  


To evaluate the prediction accuracy for learning configuration online, we rely on the widely adopted MAPE~\cite{DBLP:conf/icse/HaZ19, DBLP:journals/tosem/ChengGZ23, DBLP:conf/sigsoft/Gong023, DBLP:journals/pacmse/Gong024}. While MAPE has limitations, such as the asymmetry in penalizing overestimations and underestimations, it remains the best choice for our evaluation because it provides an intuitive measure of scale-invariant error, enabling meaningful comparisons across the systems with varying performance magnitudes~\cite{DBLP:journals/tosem/GongC25,chen2025accuracy}. 

However, since we need to measure the accuracy of predicting different batches of samples across the data stream, we extend the MAPE (denoted as mMAPE) as:
\begin{equation}  
\text{mMAPE} = \frac{1}{T} \sum_{t=1}^{T} MAPE_t  
\end{equation}  
whereby $MAPE_t$ is the MAPE measured for the $t$th batch for a total of $T$ timesteps/batches.

To ensure statistical robustness, each experiment is repeated 30 times with different random seeds. To compare the performance of multiple approaches for \textbf{RQ1} and \textbf{RQ2}, we use the Scott-Knott Effect Size Difference (Scott-Knott ESD) test \cite{scott1974cluster}, which is commonly used in software performance studies \cite{DBLP:conf/sigsoft/Gong023, chen2025accuracy, DBLP:journals/tse/GongCB25, DBLP:journals/pacmse/Gong024, DBLP:journals/tosem/GongC25} because it considers both statistical significance and practical effect sizes, avoiding misleading conclusions from minor but statistically significant differences. Scott-Knott ESD iteratively partitions approaches into distinct groups in a hierarchical manner based on effect sizes. For example, Scott-Knott ESD test might rank three approaches as two groups: $\{A,B\}$ and  $\{C\}$, such that the former is ranked 1 while the latter has 2, then we say that $A$ and $B$ are statistically similar but they are significantly better than $C$.

For pairwise comparisons in the ablation study of \textbf{RQ3}, we utilize the non-parametric Wilcoxon Signed-Rank test~\cite{DBLP:reference/stat/ReyN11} to assess whether the performance differences between \Model~and its ablated variants are statistically significant. We use a confidence level of $0.05$, i.e., the comparison is statistically significant if $p<0.05$.

\section{Evaluation}
\label{sec:evaluation}

\begin{table*}[t!]
\caption{The median and interquartile range of mMAPE, denoted as Med (IQR), for \Model~and the state-of-the-art approaches for all the subject systems over 30 runs. For each case, \setlength{\fboxsep}{1.5pt}\colorbox{beaublue!80}{blue cells} mean \Model~has the best median mMAPE; or \setlength{\fboxsep}{1.5pt}\colorbox{red!20}{red cells} otherwise. The one(s) with the best rank ($r$) from the Scott-Knott ESD test is in bold.}
\begin{adjustbox}{width=\textwidth,center}
\setlength{\tabcolsep}{0.6mm}
\begin{tabular}{lll|ll|ll|ll|ll|ll|ll|ll|ll|ll}
\toprule
\multirow{2}{*}{\textbf{System}} & \multicolumn{2}{c}{\textbf{\texttt{\Model}}} & \multicolumn{2}{c}{\textbf{\texttt{ARF}}} & \multicolumn{2}{c}{\textbf{\texttt{SPR}}} & \multicolumn{2}{c}{\textbf{\texttt{SeMPL}}} & \multicolumn{2}{c}{\textbf{\texttt{BEETLE}}} & \multicolumn{2}{c}{\textbf{\texttt{DaL}}} & \multicolumn{2}{c}{\textbf{\texttt{DaL}$_{fixed}$}} & \multicolumn{2}{c}{\textbf{\texttt{DaL}$_{\alpha}$}} & \multicolumn{2}{c}{\textbf{\texttt{RF}$_r$}} & \multicolumn{2}{c}{\textbf{\texttt{RF}$_{\alpha}$}} \\  \cline{2-21}
 & \textbf{$r$} & \textbf{Med (IQR)} & \textbf{$r$} & \textbf{Med (IQR)} & \textbf{$r$} & \textbf{Med (IQR)} & \textbf{$r$} & \textbf{Med (IQR)} & \textbf{$r$} & \textbf{Med (IQR)} & \textbf{$r$} & \textbf{Med (IQR)} & \textbf{$r$} & \textbf{Med (IQR)} & \textbf{$r$} & \textbf{Med (IQR)} & \textbf{$r$} & \textbf{Med (IQR)} & \textbf{$r$} & \textbf{Med (IQR)} \\ \hline
\textsc{SaC} & \cellcolor{beaublue!80}\textbf{1} & \cellcolor{beaublue!80}\textbf{4.36e+0 (7.74)} & 3 & 14.11e+0 (11.57) & 4 & 16.92e+0 (15.52) & 6 & 52.3e+0 (78.13) & 6 & 52.3e+0 (78.23) & 4 & 11.02e+0 (15.61) & 5 & 37.94e+0 (64.77) & 4 & 11.50e+0 (16.12) & 2 & 5.50e+0 (7.33) & 4 & 11.66e+0 (15.89) \\
\textsc{x264} & \cellcolor{beaublue!80}\textbf{1} & \cellcolor{beaublue!80}\textbf{0.32e+0 (1.05)} & 2 & 1.62e+0 (3.09) & \textbf{1} & \textbf{0.91e+0 (0.89)} & 3 & 1.66e+0 (5.01) & 3 & 2.5e+0 (4.62) & 4 & 6.39e+0 (18.97) & 4 & 4.60e+0 (14.90) & 4 & 6.44e+0 (19.13) & 4 & 6.39e+0 (18.97) & 4 & 6.45e+0 (19.14) \\
\textsc{Storm} & \cellcolor{beaublue!80}\textbf{1} & \cellcolor{beaublue!80}\textbf{27.99e+0 (60.13)} & 3 & 246.04e+0 (131.90) & 4 & 459.93e+0 (408.32) & 4 & 195.13e+0 (1157.71) & 4 & 223.09e+0 (1155.11) & 2 & 109.34e+0 (123.62) & 4 & 64.96e+0 (243.48) & 2 & 109.31e+0 (125.00) & 2 & 111.06e+0 (123.47) & 2 & 110.69e+0 (125.73) \\
\textsc{SPEAR} & \cellcolor{beaublue!80}\textbf{1} & \cellcolor{beaublue!80}\textbf{0.29e+0 (0.13)} & 2 & 1.48e+0 (0.56) & 4 & 0.87e+0 (3.61) & 2 & 1.08e+0 (1.00) & 2 & 1.2e+0 (1.08) & 2 & 1.24e+0 (1.17) & 3 & 1.27e+0 (1.86) & 2 & 1.25e+0 (1.18) & 2 & 1.23e+0 (1.17) & 2 & 1.24e+0 (1.18) \\
\textsc{SQLite} & \textbf{1} & \textbf{0.35e+0 (0.33)} & 3 & 1.43e+0 (3.72) & \cellcolor{red!20}2 & \cellcolor{red!20}0.31e+0 (0.58) & 4 & 1.77e+0 (5.81) & 4 & 3.36e+0 (6.05) & 4 & 1.88e+0 (8.20) & 5 & 1.05e+0 (9.61) & 4 & 1.90e+0 (8.28) & 4 & 1.88e+0 (8.20) & 4 & 1.91e+0 (8.28) \\
\textsc{NGINX} & 3 & 0.11e+0 (0.03) & 6 & 0.9e+0 (0.31) & 7 & 1.87e+0 (0.17) & 4 & 0.17e+0 (0.19) & 4 & 0.17e+0 (0.19) & \cellcolor{red!20}\textbf{1} & \cellcolor{red!20}\textbf{0.09e+0 (0.06)} & 5 & 0.39e+0 (0.59) & 2 & 0.10e+0 (0.06) & \cellcolor{red!20}\textbf{1} & \cellcolor{red!20}\textbf{0.09e+0 (0.06)} & 2 & 0.10e+0 (0.06) \\
\textsc{ExaStencils} & \cellcolor{beaublue!80}\textbf{1} & \cellcolor{beaublue!80}\textbf{2.64e-2 (0.00)} & 5 & 0.12e+0 (0.01) & 7 & 0.15e+0 (2.06) & 4 & 0.1e+0 (0.01) & 4 & 0.1e+0 (0.01) & 2 & 2.66e-2 (0.00) & 6 & 0.15e+0 (0.02) & 3 & 2.69e-2 (0.00) & 2 & 2.65e-2 (0.00) & 3 & 2.69e-2 (0.00) \\
\textsc{DeepArch} & \cellcolor{beaublue!80}\textbf{1} & \cellcolor{beaublue!80}\textbf{0.24e+0 (0.02)} & 3 & 0.54e+0 (0.06) & 5 & 1.01e+0 (0.30) & 4 & 0.55e+0 (0.74) & 4 & 0.55e+0 (0.74) & 2 & 0.37e+0 (0.32) & 5 & 0.78e+0 (0.61) & 2 & 0.38e+0 (0.32) & 2 & 0.37e+0 (0.32) & 2 & 0.38e+0 (0.33) \\ \hline
Average $r$ & \multicolumn{2}{l}{\cellcolor{beaublue!80}\textbf{1.25}} & \multicolumn{2}{l}{3.375} & \multicolumn{2}{l}{4} & \multicolumn{2}{l}{3.875} & \multicolumn{2}{l}{3.875} & \multicolumn{2}{l}{2.5} & \multicolumn{2}{l}{4.625} & \multicolumn{2}{l}{2.875} & \multicolumn{2}{l}{2.375} & \multicolumn{2}{l}{2.875}
\\
\bottomrule
\end{tabular}

\end{adjustbox}
\label{tb:vs_SOTA}
\end{table*}

\subsection{RQ1: \Model~against the State-of-the-art}  
\label{subsec:rq1}

\subsubsection{Method}  

{To assess the effectiveness of \Model~in online software performance learning, we compare it against several state-of-the-art approaches\footnote{Note that it has been shown that simply consider environmental features in the model is devastating~\cite{DBLP:journals/pacmse/Gong024}.} that adopt adaptive learning, online learning, meta-learning, and transfer learning:}

\begin{itemize}

   \item \textbf{Adaptive Random Forest (\texttt{ARF})} \cite{DBLP:conf/esann/GomesBFB18}: A general adaptive learning method designed for online regression tasks using an ensemble of randomized trees with adaptive mechanisms to handle concept drift.

  \item \textbf{Streaming Random Patches (\texttt{SRP})} \cite{DBLP:conf/icdm/GomesRB19}: A general online learning approach that dynamically creates and updates ensemble models using subsets of features and data patches.

  \item \textbf{Sequential Meta-Performance Learning (\texttt{SeMPL})} \cite{DBLP:journals/pacmse/Gong024}: A meta-learning approach for configuration performance learning across multiple environments in an offline setting. We adapt it by periodically retraining the model every timestep to accommodate streaming performance data. 

  \item \textbf{Bellwether Transfer Learner (\texttt{BEETLE})} \cite{DBLP:journals/tse/KrishnaNJM21}: A transfer learning method for configuration performance learning. Again, we extend it by allowing it to retrain every timestep.


   
\end{itemize}  

\noindent together with the baselines below that follow offline, periodic, or aggressive full retraining:

\begin{itemize}
    
   \item \textbf{Divide-and-Learn (\texttt{DaL})} \cite{DBLP:conf/sigsoft/Gong023}: The baseline approach that \Model~extended from (Section~\ref{sec:dal}). We compare \Model~with the naive use of \texttt{DaL} for online configuration performance learning that is completely retrained every timestep.

 \item \textbf{\texttt{DaL}$_{fixed}$}: A \texttt{DaL} that only trains on the first $50$ samples and never update.

  \item \textbf{\texttt{DaL}$_{\alpha}$}: A \texttt{DaL} that is retrained at every $\alpha$ timesteps as \Model. 

   \item \textbf{\texttt{RF}$_r$}: A Random Forest retrained at every timestep. 

  \item \textbf{\texttt{RF}$_{\alpha}$}: A Random Forest that is retrained at every $\alpha$ timesteps as \Model. 
\end{itemize}

The compared approaches all use their default settings. For \texttt{DaL}, we set $d=1$ and for \Model, we set $d=1$ and $\alpha=3$, because $d=1$ is generally recommended~\cite{DBLP:journals/tse/GongCB25} and $\alpha=3$ achieves the best trade-off between quality and time (see Section~\ref{subsec:rq4}). We also use $\alpha=3$ for \texttt{DaL}$_{\alpha}$ and \texttt{RF}$_{\alpha}$. All approaches are run under the same data stream, and we report the median and interquartile range (IQR) of the mMAPE over 30 runs, together with the ranks from the Scott-Knott ESD test. We also assess the clock time required to model adaptation per timestep.


\subsubsection{Results}

\begin{figure}[!t]
\centering
\footnotesize

\begin{subfigure}{0.49\columnwidth}
  \centering
  \includegraphics[width=\linewidth]{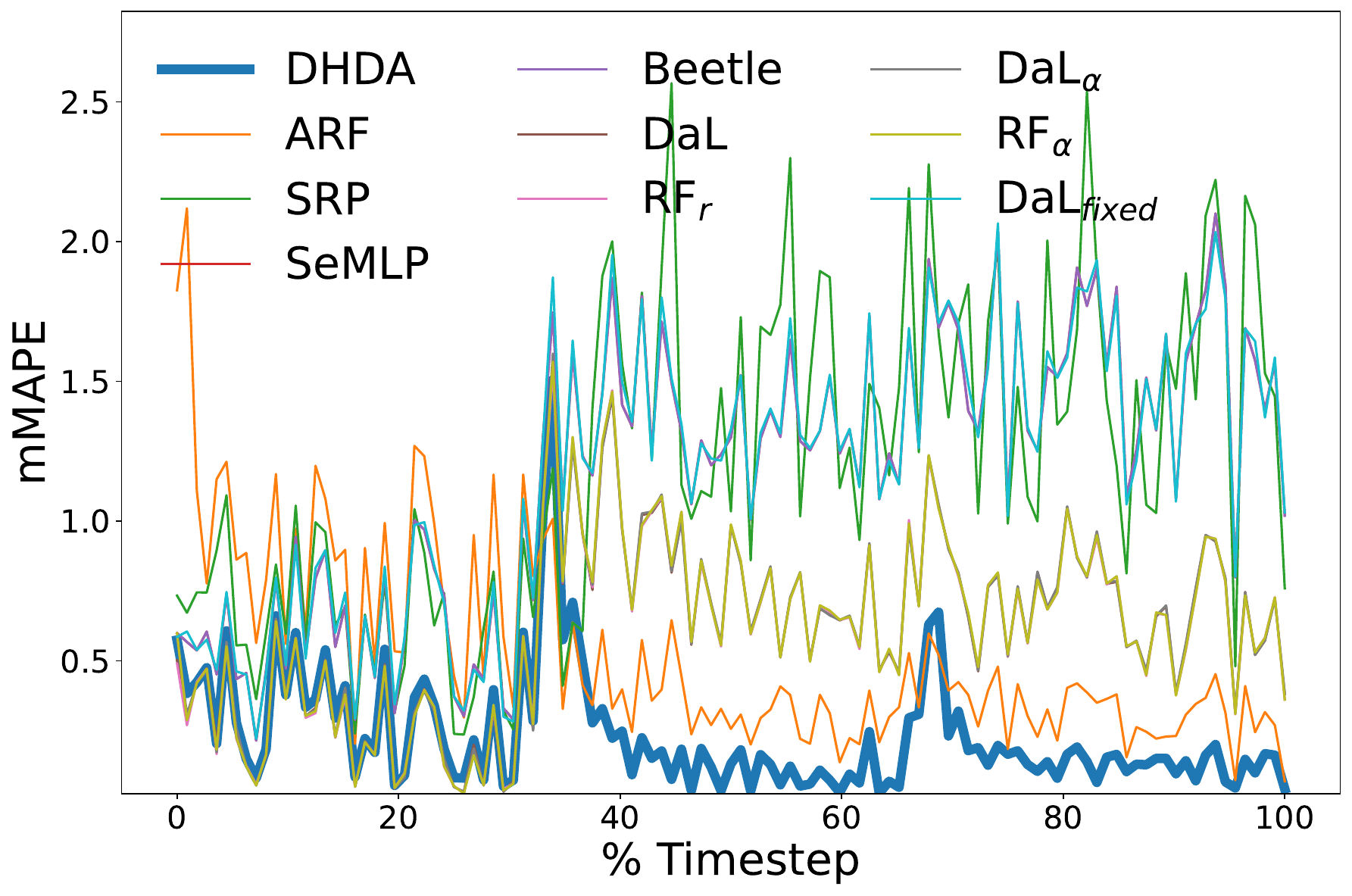} 
  \caption{\textsc{DeepArch}}
\end{subfigure}
~\hfill
\begin{subfigure}{0.49\columnwidth}
  \centering
  \includegraphics[width=\linewidth]{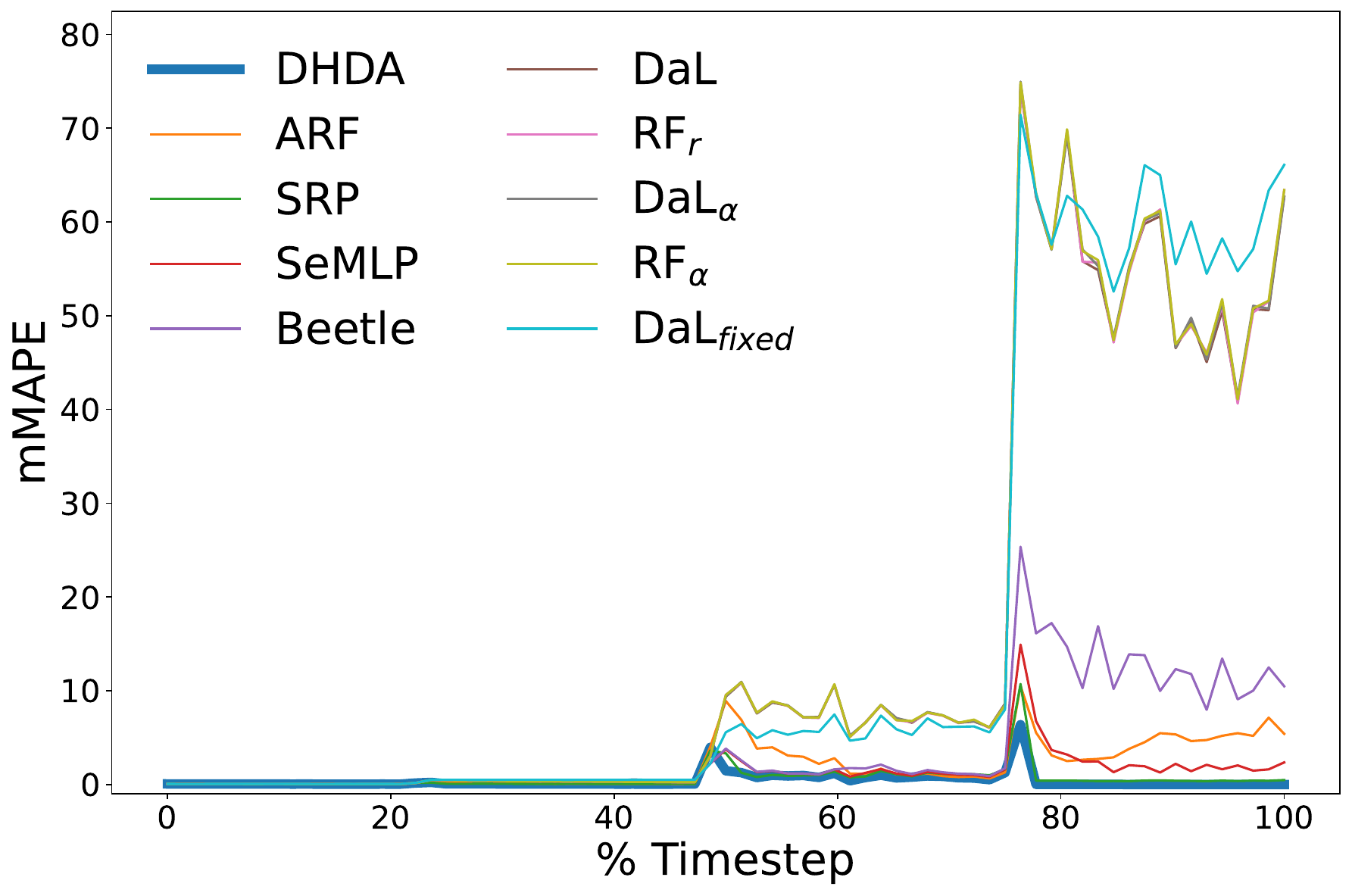} 
  \caption{\textsc{x264}}
\end{subfigure}

  \caption{Exampled accuracy traces for \textsc{DeepArch} and \textsc{x264}.}
     \label{fig:exp-trace}
\end{figure}

As can be seen from Table~\ref{tb:vs_SOTA}, \Model~achieves the best median mMAPE across 6/8 subject systems, considerably improving the others. The largest improvement is observed in \changed{\textsc{SPEAR}, where \Model~achieves a median mMAPE of 0.29, whereas the second best model, \texttt{SPR}, causes 0.87---a 2$\times$ accuracy improvement.} In terms of ranking, \Model~achieves the best rank in 7/8 systems---an average Scott-Knott rank of 1.25, which is substantially better than \changed{the best opponents \texttt{RF}$_r$ (2.375) and \texttt{DaL} (2.5)}. {To understand the stability of the results across the timesteps, Figure~\ref{fig:exp-trace} demonstrates the traces for two example systems. We see that \Model~constantly maintains a better and more stable accuracy than the others under various levels of drift.}

For the remaining two systems, \Model~is also ranked the first in \textsc{SQLite}, even though the median mMAPE is slightly worse than \texttt{SPR}. The only system where \Model~does not achieve the best median mMAPE nor is ranked the best is \textsc{NGINX}. This is attributed to \textsc{NGINX}’s relatively small magnitude of concept drifts, where periodic retraining for approaches like \texttt{DaL} is suitable.

\changed{When compared with the baselines, \Model~achieves better ranks in seven systems, while \texttt{DaL} and \texttt{RF}$_r$ are ranked the best in \textsc{NGINX}. \Model's superior performance stems from two key factors: (1) the underlying \texttt{DaL} framework handles sparse configuration data better than RF~\cite{DBLP:journals/tse/GongCB25,DBLP:conf/sigsoft/Gong023}, and (2) \Model~adapts hierarchically with proper forgetting strategies, unlike aggressive retraining where certain old data might become noise (\texttt{DaL} and \texttt{RF}$_r$) or periodic retraining that suffers from accumulated drifts between updates (\texttt{DaL}$_\alpha$ and \texttt{RF}$_\alpha$).}

\begin{figure}[t!]
  \centering
  \includegraphics[width=\columnwidth]{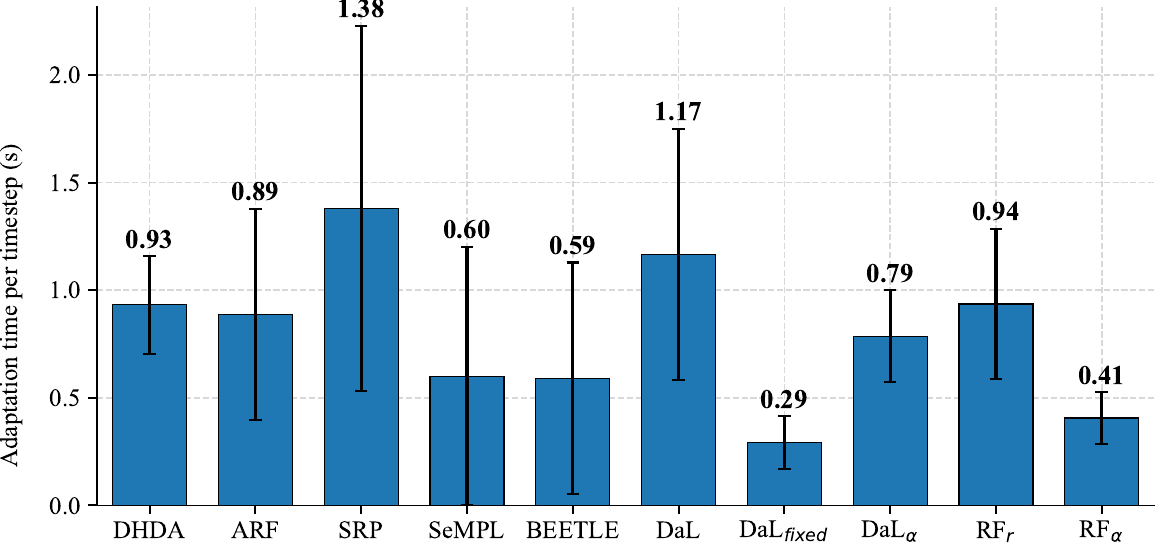}
  \caption{The mean model adaptation time and standard deviation per timestep over all systems and runs.}
  \label{fig:rq1_time}
\end{figure}

\begin{table*}[t!]
\centering
\caption{The median and IQR of mMAPE of \Model~under different local models and their counterparts paired with \Model~over 30 runs. The \colorbox{beaublue!80}{blue cells} denote the best median mMAPE. The one(s) with the best rank ($r$) from the Scott-Knott ESD test is in bold.}
\begin{adjustbox}{width=\textwidth,center}

\begin{tabular}{lll|ll|ll|ll|ll|ll|ll|ll}
\toprule
\multirow{2}{*}{\textbf{System}} & \multicolumn{2}{c}{\textbf{\Model$_{RF}$}} & \multicolumn{2}{c}{\textbf{\texttt{RF}}} & \multicolumn{2}{c}{\textbf{\Model$_{HT}$}} & \multicolumn{2}{c}{\textbf{\texttt{HT}}} & \multicolumn{2}{c}{\textbf{\Model$_{kNN}$}} & \multicolumn{2}{c}{\textbf{\texttt{$k$NN}}} & \multicolumn{2}{c}{\textbf{\Model$_{LR}$}} & \multicolumn{2}{c}{\textbf{\texttt{LR}}} \\ \cline{2-17}
 & \textbf{$r$} & \textbf{Med (IQR)} & \textbf{$r$} & \textbf{Med (IQR)} & \textbf{$r$} & \textbf{Med (IQR)} & \textbf{$r$} & \textbf{Med (IQR)} & \textbf{$r$} & \textbf{Med (IQR)} & \textbf{$r$} & \textbf{Med (IQR)} & \textbf{$r$} & \textbf{Med (IQR)} & \textbf{$r$} & \textbf{Med (IQR)} \\ \hline
\textsc{SaC} & \cellcolor{beaublue!80}\textbf{1} & \cellcolor{beaublue!80}\textbf{4.36e+0 (7.74)} & 2 & 8.95e+0 (16.49) & 3 & 9.72e+0 (16.37) & 3 & 11.38e+0 (15.65) & 2 & 5.84e+0 (8.99) & 2 & 5.14e+0 (11.35) & \textbf{1} & \textbf{4.72e+0 (7.48)} & 3 & 10.68e+0 (18.64) \\
\textsc{x264} & \textbf{1} & \textbf{0.32e+0 (1.05)} & \textbf{1} & \textbf{0.37e+0 (0.88)} & \cellcolor{beaublue!80}\textbf{1} & \cellcolor{beaublue!80}\textbf{30.60e-2 (1.03)} & \textbf{1} & \textbf{0.33e+0 (0.91)} & \textbf{1} & \textbf{30.99e-2 (1.08)} & \textbf{1} & \textbf{0.46e+0 (0.89)} & \textbf{1} & \textbf{0.70e+0 (0.26)} & \textbf{1} & \textbf{0.71e+0 (0.69)} \\
\textsc{Storm} & \cellcolor{beaublue!80}\textbf{1} & \cellcolor{beaublue!80}\textbf{27.99e+0 (60.13)} & 4 & 131.67e+0 (255.81) & 3 & 83.65e+0 (94.93) & 5 & 615.09e+0 (433.86) & 2 & 45.26e+0 (53.63) & 3 & 73.66e+0 (96.82) & 3 & 105.05e+0 (92.63) & 6 & 800.30e+0 (338.78) \\
\textsc{SPEAR} & \cellcolor{beaublue!80}\textbf{1} & \cellcolor{beaublue!80}\textbf{0.29e+0 (0.13)} & 4 & 1.13e+0 (1.02) & 2 & 0.67e+0 (0.43) & 6 & 5.85e+0 (3.56) & 2 & 0.64e+0 (0.46) & 5 & 4.18e+0 (0.88) & 3 & 0.80e+0 (0.27) & 7 & 15.20e+0 (5.02) \\
\textsc{SQLite} & 3 & 0.35e+0 (0.33) & 2 & 0.28e+0 (0.42) & \cellcolor{beaublue!80}\textbf{1} & \cellcolor{beaublue!80}\textbf{0.27e+0 (0.21)} & \textbf{1} & \textbf{0.31e+0 (0.27)} & 2 & 0.30e+0 (0.37) & \textbf{1} & \textbf{0.30e+0 (0.29)} & 4 & 0.61e+0 (0.13) & 4 & 0.47e+0 (0.26) \\
\textsc{NGINX} & \cellcolor{beaublue!80}\textbf{1} & \cellcolor{beaublue!80}\textbf{0.11e+0 (0.03)} & 4 & 0.41e+0 (0.94) & 3 & 0.59e+0 (0.22) & 7 & 2.61e+0 (1.66) & 2 & 0.42e+0 (0.1) & 5 & 1.18e+0 (0.23) & 6 & 1.66e+0 (0.09) & 8 & 5.37e+0 (0.30) \\
\textsc{ExaStencils} & \cellcolor{beaublue!80}\textbf{1} & \cellcolor{beaublue!80}\textbf{0.03e+0 (0.00)} & 2 & 0.05e+0 (0.02) & 4 & 0.09e+0 (0.01) & 6 & 0.11e+0 (0.01) & 3 & 0.08e+0 (0.00) & 5 & 0.09e+0 (0.00) & 8 & 0.62e+0 (0.01) & 7 & 0.34e+0 (0.00) \\
\textsc{DeepArch} & \cellcolor{beaublue!80}\textbf{1} & \cellcolor{beaublue!80}\textbf{0.24e+0 (0.02)} & 3 & 0.56e+0 (0.22) & 3 & 0.53e+0 (0.07) & 2 & 0.37e+0 (0.10) & 4 & 0.6e+0 (0.10) & 5 & 0.85e+0 (0.10) & 6 & 0.99e+0 (0.03) & 7 & 1.57e+0 (0.25) \\ \hline
Average $r$ & \multicolumn{2}{l}{\cellcolor{beaublue!80}\textbf{1.25}} & \multicolumn{2}{l}{2.75} & \multicolumn{2}{l}{2.5} & \multicolumn{2}{l}{3.875} & \multicolumn{2}{l}{2.25} & \multicolumn{2}{l}{3.375} & \multicolumn{2}{l}{4} & \multicolumn{2}{l}{5.375}
\\ \bottomrule
\end{tabular}

\end{adjustbox}
\label{tb:vs_local_models}
\end{table*}

Figure~\ref{fig:rq1_time} shows the model adaptation time per timestep. Notably, \Model~achieves an average of 0.93 seconds, which is highly competitive compared with the others, while maintaining the generally best accuracy across almost all systems. Overall, \changed{while \texttt{SeMPL}, \texttt{BEETLE}, \texttt{DaL}$_{fixed}$, and \texttt{RF}$_\alpha$} may be considered if lower time costs are preferred at the expense of accuracy, \Model~remains the most balanced approach for high-accuracy, real-time requirements. Thus, we say

\begin{quotebox}
   \noindent
\textit{\textbf{RQ1:} \Model~achieves statistically the best accuracy in six out of eight cases with up to \changed{$2\times$} accuracy gain, while incurring competitive model adaptation time.}


\end{quotebox}

\subsection{RQ2: \Model~under Different Local Models}
\label{subsec:rq2}

\subsubsection{Method}
Since \Model~can be paired with different local models, we evaluate its effectiveness when combined with different independent local models, including tree-based (\texttt{RF}~\cite{DBLP:journals/ml/Breiman01}, \texttt{HT}~\cite{DBLP:phd/basesearch/Tao11}), distance-based (\texttt{$k$NN}~\cite{peterson2009k}), and linear model (\texttt{LR}~\cite{DBLP:journals/technometrics/Gray02}).{ We compare those variants of \Model~with the standalone local models, each of which is incrementally updated at every timestep by refining the existing model’s sub-trees or weights with the newly arrived batch of data (after the MAPE for the new batch has been assessed)}. The evaluation follows \textbf{RQ1} with 30 runs.


\subsubsection{Results}
The results in Table~\ref{tb:vs_local_models} demonstrate that \Model~consistently improves accuracy across all local model types. For instance, when paired with \texttt{RF}, \Model~achieves an average rank of {1.25}, significantly outperforming the standalone \texttt{RF} for 1.2$\times$, which has an average rank of {2.75}. Notably, the best local model for \Model~is \texttt{RF}, which is also our default. In \textsc{NGINX}, \Model$_{\texttt{RF}}$ outperforms the second-best model \texttt{RF} for 2.7$\times$.  

In terms of subject systems, \Model$_{\texttt{RF}}$ achieves the best Scott-Knott ESD rank among seven out of eight systems. This is particularly evident in systems such as \textsc{Storm}, \textsc{SPEAR}, \textsc{NGINX}, \textsc{ExaStencils}, and \textsc{DeepArch}, where it consistently ranks as the sole best one across all other models. However, in some cases, the improvements are less pronounced. For example, in \textsc{SQLite}, \Model~with \texttt{RF} does not outperform \Model$_{\texttt{HT}}$, \texttt{HT}, and \texttt{$k$NN}. These cases could be due to the stability of performance patterns in certain systems, where frequent adaptation provides limited additional benefits. As a result, we have

\begin{quotebox}
   \noindent
   \textit{\textbf{RQ2:} \Model~is model-agnostic and consistently advances various local models with the ability to handle concept drifts, leading to better accuracy. When paired with \texttt{RF}, it achieves the best rank in seven out of eight systems, with up to 2.7$\times$ improvements.}  
   

\end{quotebox}

\begin{table*}[t!]
\centering
\caption{The median and IQR of mMAPE in the ablation study of \Model~over 30 runs. For each case, \setlength{\fboxsep}{1.5pt}\colorbox{beaublue!80}{blue cells} mean \Model~has the best median mMAPE; or \setlength{\fboxsep}{1.5pt}\colorbox{red!20}{red cells} otherwise. The $p<0.05$ cases are in bold.}
\begin{adjustbox}{width=0.7\textwidth,center}

\begin{tabular}{lll||ll||ll}
\toprule
\textbf{System} & \textbf{\Model} & \textbf{\Model$_{\texttt{NU}}$} & \textbf{\Model} & \textbf{\Model$_{\texttt{NL}}$} & \textbf{\Model} & \textbf{\Model$_{\texttt{NH}}$} \\ \hline

\textsc{SaC} & 4.36e+0 (7.74) & \cellcolor{red!20}4.30e+0 (4.65) & 4.36e+0 (7.74) & \cellcolor{red!20}3.83e+0 (6.29) & \cellcolor{beaublue!80}\textbf{4.36e+0 (7.74)} & \textbf{6.40e+0 (15.29)} \\
\textsc{x264} & \cellcolor{beaublue!80}\textbf{0.32e+0 (1.05)} & \textbf{0.37e+0 (1.10)} & \cellcolor{beaublue!80}\textbf{0.32e+0 (1.05)} & \textbf{1.10e+0 (2.00)} & \cellcolor{beaublue!80}\textbf{0.32e+0 (1.05)} & \textbf{0.55e+0 (1.16)} \\
\textsc{Storm} & \cellcolor{beaublue!80}27.99e+0 (60.13) & 29.98e+0 (62.87) & \cellcolor{beaublue!80}27.99e+0 (60.13) & 44.61e+0 (45.40) & \cellcolor{beaublue!80}\textbf{27.99e+0 (60.13)} & \textbf{99.91e+0 (244.14)} \\
\textsc{SPEAR} & \textbf{0.29e+0 (0.13)} & \cellcolor{red!20}\textbf{0.28e+0 (0.15)} & \cellcolor{beaublue!80}\textbf{0.29e+0 (0.13)} & \textbf{0.57e+0 (0.29)} & \cellcolor{beaublue!80}\textbf{0.29e+0 (0.13)} & \textbf{1.65e+0 (2.39)} \\
\textsc{SQLite} & \cellcolor{beaublue!80}\textbf{0.35e+0 (0.33)} & \textbf{0.36e+0 (0.32)} & \cellcolor{beaublue!80}\textbf{0.35e+0 (0.33)} & \textbf{0.78e+0 (1.73)} & \cellcolor{beaublue!80}0.35e+0 (0.33) & 0.37e+0 (0.66) \\
\textsc{NGINX} & \cellcolor{beaublue!80}0.11e+0 (0.03) & 0.12e+0 (0.06) & \textbf{0.11e+0 (0.03)} & \cellcolor{red!20}\textbf{0.07e+0 (0.02)} & \cellcolor{beaublue!80}\textbf{0.11e+0 (0.03)} & \textbf{0.83e+0 (1.33)} \\
\textsc{ExaStencils} & \cellcolor{beaublue!80}\textbf{2.64e-2 (0.00)} & \textbf{2.65e-2 (0.00)} & \cellcolor{beaublue!80}\textbf{0.03e+0 (0.00)} & \textbf{0.05e+0 (0.00)} & \cellcolor{beaublue!80}\textbf{0.03e+0 (0.00)} & \textbf{0.14e+0 (0.05)} \\
\textsc{DeepArch} & \cellcolor{beaublue!80}0.24e+0 (0.02) & 0.24e+0 (0.02) & \cellcolor{beaublue!80}\textbf{0.24e+0 (0.02)} & \textbf{0.36e+0 (0.03)} & \cellcolor{beaublue!80}\textbf{0.24e+0 (0.02)} & \textbf{0.95e+0 (0.25)}
\\
\bottomrule
\end{tabular}

\end{adjustbox}
\label{tb:vs_components}
\end{table*}

\subsection{RQ3: Effectiveness of the Key Components}  
\label{subsec:rq3}  

\subsubsection{Method}  

To assess the robustness of \Model, we conduct an ablation study by disabling key components and measuring their impact on accuracy. Specifically, we evaluate four ablated versions of \Model, each with one key component removed:  

\begin{itemize}  
    \item \textbf{\Model$_{\texttt{NU}}$ (No upper level online learning):} Remove upper-level global drift detection and adaptation, restricting \Model~to only react to local drifts. 
    

    \item \textbf{\Model$_{\texttt{NL}}$ (No lower level online learning):} Disable lower-level drift detection, forcing the model to rely solely on global drift detection/adaptation. 

    \item \textbf{\Model$_{\texttt{NH}}$ (No hybrid maintaining):} The models are not updated unless there is a concept drift.
    
\end{itemize}  

Each ablation variant is evaluated using the same setup as in \textbf{RQ1} over 30 runs. Yet, since we are interested in pair-wise comparisons and the data stream sequence used in the updating is unchanged, we use Wilcoxon Signed-Rank test to detect less biased statistical significance.

\subsubsection{Results}  

Results in Table~\ref{tb:vs_components} highlight the necessity of each key component in \Model. In most of the systems, \Model~consistently achieves the better median mMAPE with statistical significance (\( p < 0.05 \)), confirming that removing any component degrades performance.  

\Model$_{\texttt{NU}}$ (without upper level learning) performs worse than \Model~in median mMAPE for six out of eight systems, four of which show statistical significance. For the only cases where \Model$_{\texttt{NU}}$ performs better (\textsc{SPEAR}), the reason is that the sudden global drifts rarely occur, and hence local drift detection/adaptation is sufficient.

Subsequently, \Model$_{\texttt{NL}}$ (without lower level learning) exhibits degradation in mMAPE for six systems, while five of them show \( p<0.05 \). In particular, in \textsc{x264}, removing local drift adaptation severely impacts prediction accuracy, causing the median mMAPE to worsen from 0.32 in \Model~to 1.10 in \Model$_{\texttt{NL}}$---a degradation by a factor of 2.44.  These results indicate that without local adaptation, \Model$_{\texttt{NL}}$~fails to respond rapidly to frequent small-scale shifts, degrading overall prediction accuracy. Yet, for \texttt{NGINX}, \Model~is worse because although \texttt{NGINX} mostly suffers local drifts, the magnitude is however relatively small, hence the lower level drift adaptation in \Model~does not bring many benefits but incurs disruption.




Finally, \Model$_{\texttt{NH}}$ (without hybrid maintaining) suffers severe accuracy loss in all eight systems ($p<0.05$ in seven). These results emphasize that hybrid model maintenance plays a crucial role in keeping local models stable even in cases when no concept drift is detected. Overall, we conclude that

\begin{quotebox}
   \noindent
\textit{\textbf{RQ3:} The key components in \Model~can significantly improve the accuracy in general, or at least do not harm the performance, especially when a certain level of concept drifts occur frequently.}  
\end{quotebox}

\subsection{Sensitivity to the Parameter $\alpha$}  
\label{subsec:rq4}  

\subsubsection{Method}  

The parameter \( \alpha \) in \Model~determines how frequently local models undergo full retraining in the hybrid model maintenance process. To understand the sensitivity of \Model~to $\alpha$, we test \Model~under different \( \alpha \) settings, ranging from rather infrequent retraining (\( \alpha = 7 \)) to full retraining (\( \alpha = 1 \)). We report on both the mMAPE and the clock time required to model adaptation per timestep. Other settings are the same as \textbf{RQ1}.

\subsubsection{Results}
\begin{figure}[!t]
\centering
\footnotesize

\begin{subfigure}{0.49\columnwidth}
  \centering
  \includegraphics[width=\columnwidth]{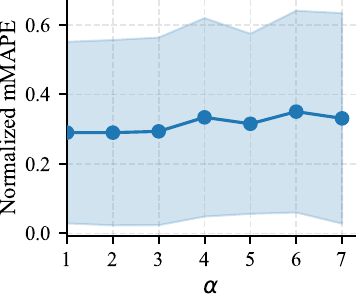} 
  \caption{mMAPE}
\end{subfigure}
~\hfill
\begin{subfigure}{0.49\columnwidth}
  \centering
  \includegraphics[width=\columnwidth]{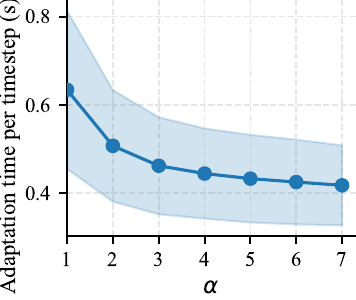} 
  \caption{Clock time}
\end{subfigure}

  \caption{The mean mMAPE/model adaptation time and its standard deviation showing the sensitivity of \Model~to the parameter $\alpha$ over all systems and runs.}
    \label{fig:sensitivity}
\end{figure}

Figure~\ref{fig:sensitivity} presents the effect of the parameter \( \alpha \) on both mMAPE and clock time for model adaptation. From Figure~\ref{fig:sensitivity}a, we observe that the mean mMAPE remains relatively stable when $1\leq \alpha \leq3$, even though a bigger $\alpha$ implies less frequent retraining, hence the correlation between old and new data samples is learned less. In contrast, the accuracy starts to drop when $\alpha\geq4$, suggesting that this is the point where only relying on incremental update becomes harmful. On the other hand, Figure~\ref{fig:sensitivity}b reveals a clear decreasing trend in model adaptation time as \( \alpha \) increases, which is as expected, and we see that when $\alpha=3$ the time reduction becomes much slower. This demonstrates that reducing retraining frequency significantly improves computational efficiency, lowering the overall adaptation cost. As a result, we set $\alpha = 3$ as the default, serving as a reasonable trade-off between accuracy and overhead. Therefore, we say that

\begin{quotebox}
   \noindent
\textit{\textbf{RQ4:} The accuracy of \Model~is sensitive to $\alpha$ when it is big enough; in contrast, $\alpha$ is negatively and monotonically correlated with model adaptation time. Yet, setting $\alpha = 3$ leads to a reasonable trade-off.}


\end{quotebox}

\section{Discussion}  
\label{sec:discussion}  

\subsection{Why does \Model~Work?}

\begin{figure}[!t]
\centering
\footnotesize

\begin{subfigure}{0.49\columnwidth}
  \centering

    \includegraphics[width=\linewidth]{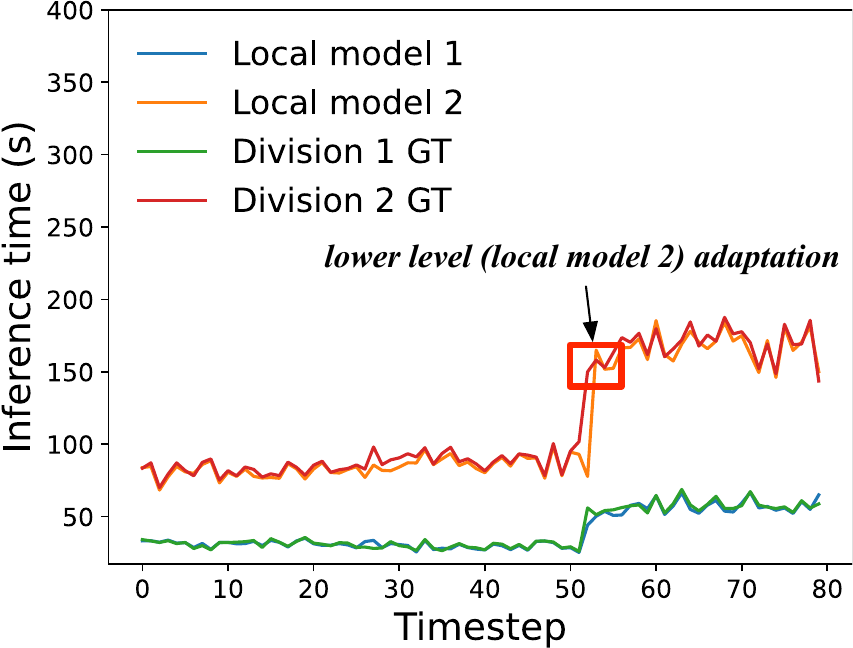} 
  
   \caption{\textsc{DeepArch}}
\end{subfigure}
~\hfill
\begin{subfigure}{0.49\columnwidth}
  \centering
      \includegraphics[width=\linewidth]{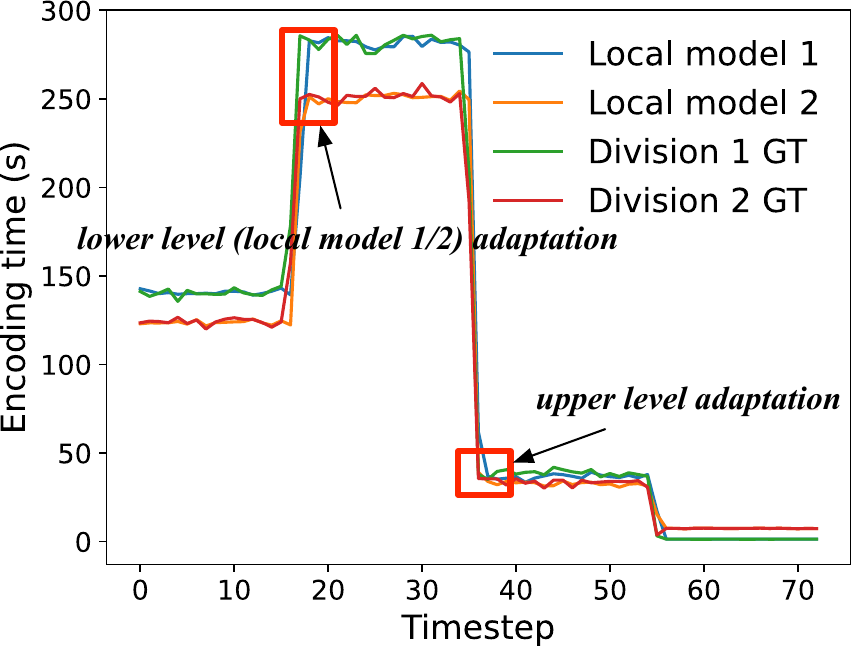} 
 \caption{\textsc{x264}}
\end{subfigure}

  \caption{Examples of performance prediction traces of all divisions in \Model. GT denotes ground truth.}
     \label{fig:discussion_example}
\end{figure}

To understand why \Model~work, Figure~\ref{fig:discussion_example} shows the common performance prediction traces within each division of \Model~for two systems. We can see that, for Figure~\ref{fig:discussion_example}a, \Model~is able to adapt to local drift that occurred only within one of the divisions, while the other model is unchanged and can still maintain good predictions. In contrast, for Figure~\ref{fig:discussion_example}b, \Model~adapts at the lower level on both local models initially, but later on the concept drifts become much more globally harmful, \Model~then adapts at the upper level, redividing the samples and retraining all local models again. These examples evidence the superiority of \Model~which is consistent with our theory presented in Section~\ref{sec:pre}.


\subsection{Is \Model~Practically Useful?}  

{Being a model, \Model~cannot make decisions, but it serves as an important foundation for such when paired with heuristic/deterministic algorithms~\cite{DBLP:journals/tosem/ChenL23}. For example, in configuration tuning with Bayesian optimization, where a single configuration measurement might take up to hours~\cite{DBLP:journals/pacmse/0001L24,DBLP:journals/tosem/ChenL23a}, \Model~can serve as the surrogate model that cheaply evaluates the configurations' performance~\cite{DBLP:conf/icse/ChenChen26}. It also allows rapid updates/adaptations to new data via learning online. For self-adaptation, a model, which can be updated as the system runs, can help assess system performance at runtime without affecting the system, e.g., model-assisted genetic-algorithm planner~\cite{DBLP:journals/tosem/ChenLBY18,DBLP:conf/icse/YeChen25,DBLP:conf/wcre/Chen22}.} Yet another example is that \Model~can also help with prioritization of configuration performance testing~\cite{DBLP:conf/icse/MaChen25}, e.g., we can use \Model~to evaluate the configuration, rank them, and choose the one with worse performance to measure/test first, which increases the likelihood of finding performance bugs earlier.

A related challenging question to answer for the above is whether the accuracy improvements observed from \Model~is practically beneficial~\cite{chen2025accuracy}. Indeed, the applications of the approach can be vast, and it is almost impossible to foresee how the resulting model can be used. Yet, to provide insights on this aspect, we refer to the data collected by Chen et al.~\cite{chen2025accuracy} from a large-scale empirical study, which provides ``hints'' on how much accuracy improvement tends to be practically useful.

Chen et al.~\cite{chen2025accuracy} summarize the average extents of accuracy (MAPE) improvement of an approach $A$ over the other $B$ can lead to statistically significant benefits when being used to guide a model-based configuration tuning. For example, if $B$ has an MAPE of $0.15$ ($15\% \in [20\%,30\%]$), then $A$ needs to reduce that by at least the threshold of $0.093$ for one to observe statistically significant improvements from the tuning when using $A$ instead of $B$.

\begin{table}[t]
\caption{The median accuracy improvements of \Model~over the second best approach that can lead to statistically significant benefits in practical use of the resulted model. We consider only the systems that \Model~is ranked the best. Full data can be accessed here: \textcolor{blue}{\texttt{\url{https://github.com/ideas-labo/model-impact/blob/main/RQ_supplementary/RQ5/supplement.pdf}}}.}
\label{tab:useful}
\begin{adjustbox}{width=\linewidth, center}
  \centering
  \begin{tabular}{lllll}
    \toprule
     \textbf{System}&
    \textbf{Second Best} & \textbf{$\Delta$ vs. \Model} & \textbf{The Least $\Delta$~\cite{chen2025accuracy}}&\textbf{Useful?}\\  \midrule

\textsc{SaC}&\changed{\texttt{RF}$_{r}$}&1.14e+0&0.68e+0&\textcolor{teal}{\ding{52}}\\
\textsc{x264}&\texttt{SPR}&0.59e+0&0.35e+0&\textcolor{teal}{\ding{52}}\\
\textsc{Storm}&\changed{\texttt{DaL}$_{fixed}$}&36.97e+0&0.68e+0&\textcolor{teal}{\ding{52}}\\
\textsc{SPEAR}&\texttt{SPR}&0.58e+0&0.38e+0&\textcolor{teal}{\ding{52}}\\
\textsc{ExaStencils}&\changed{\texttt{RF}$_{r}$}&0.01e-2&0.07e-1&\textcolor{red}{\ding{55}}\\
\textsc{DeepArch}&\texttt{DaL}&0.13e+0&0.13e+0&\textcolor{teal}{\ding{52}}\\

    \bottomrule
  \end{tabular}
\end{adjustbox}
\end{table}

Drawing on the above, we showcase the systems where \Model~performs the best and compare the accuracy improvement to the second best approach. As we can see from Table~\ref{tab:useful}, clearly, the accuracy improvement made by \Model~over the second best approach has exceeded the $\Delta$ thresholds on five out of six cases, meaning that, in general, the improved accuracy from \Model~can be useful and beneficial for practical model-based configuration tuning.

\subsection{Assumptions and Limitations}
\label{subsec:assumptions_limitations}

{A key assumption of \Model~is that the data from different environments have uncertain/varying correlations, i.e., data from one environment cannot always be useless (suitable for incremental learning), nor consistently useful (suitable for simple retraining), for the others. There would be some uncertain dynamics involved, hence the dynamic adaptation in \Model~is meaningful. Prior studies have revealed those for configuration data~\cite{DBLP:conf/kbse/JamshidiSVKPA17, muhlbauer2023analysing}.}

{Several limitations should be acknowledged. (1) \Model~is less effective for abrupt changes, e.g., rapid migration between completely different hardware, as the drift detection/adaptation needs time to reflect; (2) \Model~cannot cold start, i.e., it requires some accumulated data for the drift detector to work properly; (3) Although we found $\alpha=3$ is generally reasonable, it might not be universally optimal.}

\subsection{Threats to Validity}  
\label{sec:threats}  

\textbf{Internal Threats.}  
Internal threats to validity are related to the parameter settings used in \Model~and its baseline comparisons. To ensure fairness, we strictly follow their original parameter settings and ensure their effectiveness in the online setting. We also conducted a sensitivity analysis on the parameter $\alpha$ across multiple subject systems and found that $\alpha=3$ is a reasonable trade-off between accuracy and time. Yet, we agree that those parameter settings might not be optimal for all scenarios.



\textbf{Construct Threats.}  
Threats to construct validity may lie in the choice of evaluation metrics. We use MAPE as the primary metric due to its wide adoption in software performance learning \cite{DBLP:journals/tse/KrishnaNJM21, DBLP:journals/pacmse/Gong024}. However, MAPE is sensitive to small actual values and penalizes over- and under-estimations asymmetrically. To mitigate this, we conduct statistical validation using the Scott-Knott ESD test and Wilcoxon Signed-Rank test for 30 runs.  

\textbf{External Threats.}  
External validity could be raised from the subject systems used. While we use eight commonly studied configurable systems from diverse domains \cite{DBLP:conf/sigsoft/JamshidiVKS18, DBLP:journals/tse/KrishnaNJM21, DBLP:conf/icse/webertwins}. We have also assessed \Model~paired with different local models. Indeed, evaluating more systems might further consolidate our results.


\section{Related Work}  
\label{sec:related_work}  

\textbf{Single-environment Offline Configuration Performance Model.} Machine learning approaches have been widely applied to predict configuration performance in a single static environment. \texttt{SPLConqueror} \cite{DBLP:journals/sqj/SiegmundRKKAS12} applies a performance-influence model based on linear regression to capture the impact of configuration options. \texttt{DECART} \cite{DBLP:journals/ese/GuoYSASVCWY18} extends \CART~with a novel sampling technique to improve performance learning efficiency. Deep learning-based methods have also been explored, such as \texttt{DeepPerf} \cite{DBLP:conf/icse/HaZ19}, which leverages regularized deep neural networks to learn the sparse interactions between options, and \texttt{HINNPerf} \cite{DBLP:journals/tosem/ChengGZ23}, which employs hierarchical interaction neural networks to capture the hierarchical dependencies between configurations and performance. \texttt{DaL} \cite{DBLP:conf/sigsoft/Gong023} further improves the accuracy level by handling both feature and data sparsity. 

However, these approaches assume an overly optimistic static environment and do not consider the evolving performance landscapes in dynamic software systems. 

\textbf{Multi-environment Offline Configuration Performance Model.} To extend performance prediction across multiple environments, meta-learning and transfer learning techniques have been introduced. For example, transfer learning methods such as \texttt{BEETLE} \cite{DBLP:journals/tse/KrishnaNJM21} and the other similar approach \texttt{tEAMS} \cite{DBLP:journals/tse/MartinAPLJK22} aim to transfer knowledge from one environment to another, improving model adaptability. \texttt{SeMPL} \cite{DBLP:journals/pacmse/Gong024} further employs historical knowledge from multiple environments in sequential meta-learning to better learn configuration performance. 

While these methods enhance generalization across environments, they rely on offline training, requiring documented data from known environments before making predictions, which limits their applicability in real-time evolving settings.  

\textbf{General Online Learning Models.} Online learning models exist for handling concept drifts in different forms~\cite{DBLP:conf/esann/GomesBFB18,DBLP:conf/icdm/GomesRB19,DBLP:conf/ida/BifetG09,DBLP:journals/eswa/SoaresA15,DBLP:journals/tnn/ElwellP11}. Among others, \texttt{ARF}~\cite{DBLP:conf/esann/GomesBFB18} uses dedicated detectors to monitor data distribution changes and trigger model updates when drift is identified. Examples include Adaptive Random Forest, which integrates drift detectors into an ensemble framework. \texttt{SRP} \cite{DBLP:conf/icdm/GomesRB19} is a framework that dynamically updates model components using randomly selected features and data subsets, and Hoeffding Adaptive Tree (\texttt{HAT}) \cite{DBLP:conf/ida/BifetG09} is another similar model that applies the Hoeffding bound to detect significant changes in the feature distribution.

However, unlike \Model, those approaches cannot handle the unique sparsity and the inherited hierarchical concept drifts presented in configuration data. Further, they have ignored the expensive measurements of configuration data; hence, they do not work well in small datasets like configuration data.

\section{Conclusion}  
\label{sec:conclusion}  

This paper presents \Model, a framework for online configuration performance learning, handling the unique global and local drifts presented in configuration data. The key novelty of \Model~is that, relying on the two level structure from \texttt{DaL}, it performs dually hierarchical drift adaptation: at the upper level, the configuration data can be re-divided and all local model is retained only when needed to handle global drifts; at the lower level, the local model of each division can detect and adapt to local drifts. By evaluating \Model~on eight systems and against five baseline/state-of-the-art, we demonstrate that \Model:

\begin{itemize}
    \item considerably improves the accuracy against concept drifts in 6 out of 8 systems with up to \changed{$2\times$} improvements;
    \item can enhance the concept drift adaptation ability of various local models;
    \item and leads to competitive model adaptation time in the magnitude of seconds.
\end{itemize}

Future work along this line of research can be fruitful, including dynamically reusing old local models and more advanced estimation of concept drift types.


\section*{Acknowledgment}
This work was supported by a NSFC Grant (62372084) and a UKRI Grant (10054084).

\balance
\bibliographystyle{ACM-Reference-Format}
\bibliography{references}

\end{document}